\documentclass[twocolumn,superscriptaddress, preprintnumbers,pra]{revtex4}
\usepackage{graphicx}
\usepackage{dcolumn}
\usepackage{amssymb}
\usepackage{amsmath}
\usepackage{bbm}
\usepackage{setspace}
\usepackage{bm}
\usepackage{hyperref}

\hypersetup
{
colorlinks=true, 
citecolor=blue
}
\begin{document}

\title{High-order harmonic generations in tilted Weyl semimetals}
\author{Zi-Yuan Li}
\email{lizy@aircas.ac.cn}
\author{Qi Li}
\author{Zhou Li}
\affiliation{GBA Branch of Aerospace Information Research Institute, Chinese Academy of Sciences, Guangzhou 510700, China}

\begin{abstract}
We investigate high-order harmonic generations (HHGs) under the comparison of Weyl cones in two types. Due to the hyperboloidal electron pocket structure, strong noncentrosymmetrical generations in high orders are observed around a single type-II Weyl point, especially at frequency zero.
Such remarkable DC signal is proved to have attributions from the intraband transition after spectral decomposition.
Under weak pulse electric field , the linear optical response of a non-tilted Weyl cone is consistent with the  Kubo theory.
With more numerical simulations, we conclude the non-zero chemical potential can enhance the even-order generations, from the slightly tilted system to the over-tilted systems. In consideration of dynamical symmetries, type-I and -II Weyl cones also show different selective responses under the circularly polarized light.
Finally, using a more realistic model containing two pairs of Weyl points, we demonstrate the paired Weyl points with opposite chirality could suppress the overall even-order generations.
\end{abstract}
\date{today}
\pacs{Valid PACS appear here}
\maketitle
\section{introduction}
The study of nonlinear optics is mainly accelerated by the interaction of intense laser field with materials \cite{boyd2020}. As an up-conversion process of photons, high-order harmonic generations (HHGs) represent the extreme nonlinear electron dynamic process interacting with light. The most early non-perturbative HHGs was found in the rare gases \cite{mcpherson1987} and the dynamics was explained by the semiclassical three-step model \cite{krause1992,schafer1993,corkum1993,ghimire2019}.
The discovery of nonlinear generations in solid-state materials \cite{ghimire2011} has attracted people's attention on attosecond pulse generation \cite{vampa2017,garg2018,li2020} and its potential applications in nano-scale devices \cite{sivis2013,han2016,vampa2017plasmon}.
HHGs can capture the dynamic information of electrons, which makes it a useful tool for reconstructing the energy band structures \cite{vampa2015all,tancogne2017,lanin2017,li2020determination}, exploring topological phase transitions and symmetry breakings \cite{silva2019,chacon2020,schmid2021,bai2021}.
The microscopic mechanisms, such as the relationship between cut-off energy and driver laser wavelength, the coupling dynamics of optical inter- and intraband excitation are also being solved \cite{golde2008,vampa2014theoretical,vampa2015semiclassical,ikemachi2017trajectory,tancogne2017impact,floss2018,sato2021high}.

The superior properties of semimetals, such as high carrier mobility, low charge carrier density, low effective mass, and large magnetoresistance bring promising prospect to novel devices \cite{armitage2018,hu2019}. As one of the most representative Dirac semimetals (DSMs), graphene shows high conversion efficiencies of terahertz high harmonics \cite{mics2015,yoshikawa2017,hafez2018}, that motivates people to explore HHGs' control strategy associated with stacked, twisted structures or heterostructures in graphene, graphene-like two dimensional materials \cite{shan2018,tancogne2018,ikeda2020high,alonso2021giant}. Recently HHGs in the 3D DSM $\rm{Cd_{3}As_{2}}$ is observed at room temperature \cite{cheng2020}, which considered to be originated from coherent intraband acceleration of Dirac electrons.
The quasi-particles in these materials are governed by the Dirac or Dirac-like equations.
By breaking either the time-reversal symmetry or inversion symmetry, a single Dirac node can be separated into two Weyl nodes with opposite chirality. Surface Fermi arcs that connect the pair of Weyl points (WPs) are predicted and observed both in material calculations \cite{huang2015,weng2015,murakami2017} and experiments \cite{lv2015,lv2015observ,xu2015,arnold2016,xu2015discovery,xu2016observation}, like TaAs, TaP and other Weyl semimetals (WSMs).
While the potential energy component of $E(\bm{k})$ is larger than
the kinetic component at some $\bm{k}$ points, the Weyl cone will be over tilted, with the hyperboloidal electron and hole pockets intersecting at the nodal point. We call it the type-II semimetal \cite{soluyanov2015}.
Owing to the linear energy dispersion and large Berry curvature dipole near the WPs, WSMs are expected to perform strong nonlinear responses \cite{sodemann2015,zhang2018,rostami2018,kang2019,wawrzik2021}.
The strong photovoltaic process is predicted with the combination of inversion symmetry breaking and tilts of the Weyl cone \cite{chan2017}. Up to now, giant nonlinear optical responses are found in both types of WSMs with broken inversion symmetry, such as the second-harmonic generations in TaAs \cite{sirica2019,sirica2022},
photocurrent from the circular photogalvanic effect in type-II WSMs TaIrTe4, WTe2 \cite{de2017,wang2019,ma2019}, and high harmonics from WSM $\rm{WP_{2}}$ \cite{lv2021}.

Our work focuses on the influence of tilted structures to the high-order harmonics in Weyl semimetal. Numerical evolution of the time-dependent Dirac equation (TDDE) is performed to simulate the electron dynamics with laser \cite{ishikawa2010,lim2020,ikeda2020high}.
In the type-II cones, the electron pocket is occupied, but the hole pocket is unoccupied. The electrons stimulated from such asymmetrical structures exhibit tremendous large direct current (DC) signal, even- and odd-order high harmonic generations. Through the diagonal transformation, we figure out the roles of
intraband and interband transitions in the HHGs.
Generations in zero and non-zero chemical potential cases are all considered.
We also reveal the selective generations are broken in the over-tilted Weyl cone with a circularly polarized light. In the last part, the nonlinear optical generations are investigated in a lattice model existing two pairs of type-II Weyl cones. The sign of chirality determines the interference effects of the harmonics around the cones.

\section{hamiltonian and electron dynamics coupling with laser}
\begin{figure}[h]
\centering
\includegraphics[width=1\linewidth]{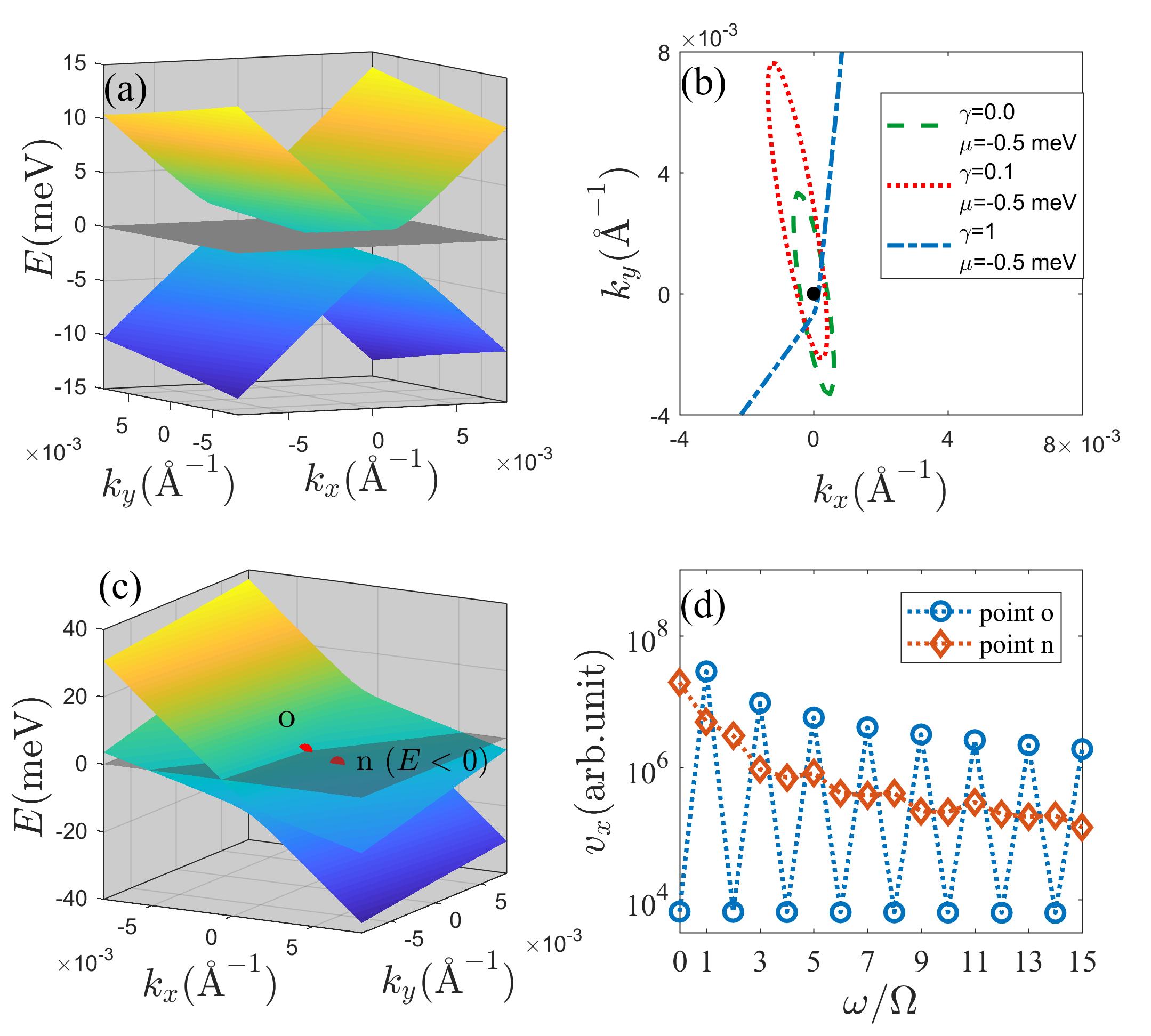}
\caption{Schematic of the band structure for (a) non-tilted and (c) over-tilted cones,
which refer to the hamiltonian in Eq.~(\ref{eq:one}) with
$\gamma=0$ and $\gamma=1$ respectively.
(b) The valence-band contours at chemical potential $\mu=$ -0.5 meV are plotted where the dashed line, dotted line and dot-dashed line correspond to the the cones with different tilt parameters $\gamma=0, 0.1, 1$ respectively. (d) shows the Fourier harmonics of $v_{x}(t)$'s oscillating part at the original point o and one point n on the upper band of the electron pocket, which both marked in (c).}
\label{Fig1}
\end{figure}

 The Weyl points (WPs) in WSMs with opposite topological charge are usually not degenerate in energy. Thus it is possible that only one node near the chemical potential is stimulated while the another one is Pauli-blocked at sufficiently low energy.
 To demonstrate the optical responses around the type-I and -II cones,
we adopt the continuous model of one Weyl point taken from Ref.~\cite{soluyanov2015}
 \begin{eqnarray}
 \label{eq:one}
&&\hat{H}(\bm{k})=  \nonumber \\
&&\gamma (Ak_{x}+Bk_{y})+(ak_{x}+ck_{y})\hat{\sigma}_{y}+(bk_{x}+dk_{y})\hat{\sigma}_{z}+ek_{z}\hat{\sigma}_{x}  \notag \\
 \end{eqnarray}
where $\hat{\sigma}_{x, y, z}$ are Pauli matrices and the parameters are $A=-2.738; B=0.612; a=0.987; b=1.107; c=0.0; d=0.270$ and $e=0.184$ with units [$\rm{eV\AA}$]. We multiply $\gamma$ on the scalar terms to control how tilted the cone is.
As $\gamma$ ranges from 0 to 1, the cone will be tilted from type-I to -II, shown in Fig.~\ref{Fig1} (a)(c) and it starts tangent with the energy surface which passes through the node point, at critical value $\gamma_{c}\approx 0.173$.

The eigenvalues also contain the tilted and non-tilted parts
\begin{eqnarray}
E_{\pm}(\bm{k})=E_{{\rm{t}}}\pm E_{{\rm{nt}}}
\label{eigenstates}
\end{eqnarray}
specifically that
$E_{{\rm{t}}}=\gamma(Ak_{x}+Bk_{y})$, $E_{{\rm{nt}}}=\sqrt{e^{2}k_{z}^{2}+(bk_{x}+dk_{y})^{2}+(ak_{x}+ck_{y})^{2}}$,
and signs $\pm$ refer to the upper and lower bands.

The characteristic difference between the two types of Weyl cones is the pockets' structure as illustrated in Fig.~\ref{Fig1}(b).  For the non-tilted cone ($\gamma=0$), the edge of the contours stays elliptical and centrosymmetric. While $0<\gamma<\gamma_{c}$, it starts to be distorted and uncentrosymmetric but still enclosed. When the tilted parameter is over the critical value $\gamma_{c}$, the electron and hole pockets of the type-II cones are formed, in which
the boundary of pockets is hyperboloidal and the area of pockets goes larger with the $\gamma$ increasing.

To see how the electrons are driven by the laser under such different structures of Weyl cones, we implant
the vector potential $\bm{A}(t)$ of the pulse's electric filed into the electron's momentum terms.
Therefore the Hamiltonian in time is given by \cite{ishikawa2010}
\begin{equation}
\label{eq:four}
\hat{H}(\bm{k},t)=\hat{H}(\bm{k}+e_{l}\bm{A}(t)),
\end{equation}
where $e_{l}$ is the elementary charge.
In particular, the vector potential's form is
\begin{equation}
\label{eq:vp}
A_{j}(t)=\frac{E_{0}}{\Omega}\exp[-2\ln 2(\frac{t}{{t_{\rm{0}}}})^{2}]\sin(\Omega t)
\end{equation}
where $j$ is the polarized direction, $E_{0}$ is the peak amplitude of the electric field, $\Omega$ is the photon's frequency and $t_{0}$ in the Gaussian envelope function sets to be $5T$ with $T=2\pi/\Omega$.

The velocity components from both bands are given by
$v_{x}=\gamma A\pm [(a^{2}+b^{2})k_{x}+(ac+bd)k_{y}]/L$,
$v_{y}=\gamma B\pm [(ac+bd)k_{x}+(c^{2}+d^{2})k_{y}]/L$ and
$v_{z}=\pm e^{2}k_{z}/L$,
with $L=\sqrt{(ak_{x}+ck_{y})^{2}+(bk_{x}+dk_{y})^{2}+e^{2}k_{z}^{2}}$.
Firstly we consider the quasiclassical kinetic approach by
replacing the momentum $\bm{k}$ inside the velocities with $\bm{k}+e_l\bm{A}(t)$.  So we can describe the electrons' movements by $\bm{v}(t)$ inside each band.
Typically the oscillation happens around the original point o is
\begin{eqnarray}
v_{x}(t)=\gamma A \pm {\rm{sgn}}(A_{x}(t))\sqrt{a^{2}+b^{2}}
\end{eqnarray}
where the electric field is chosen to be linearly polarized along x direction and the tilted term $\gamma A$ contributes only as constant in the velocities. Considering the unique pocket structure in type-II Weyl cone, we also check the upper-band oscillation at point n residing in the electron pocket ($\bm{k}=[0.002,0,0]\rm{\AA}^{-1}$).
Since term $\gamma A$ contributes the same direct signal strength in different $\bm{k}$, we only discuss the Fourier harmonics of $v_{x}(t)$'s oscillating part.
It is shown in Fig.~\ref{Fig1}(d) that point o contains all odd Fourier harmonics and the amplitudes fall down slowly.
Meanwhile at point n, the electron inside the pocket goes through the Weyl point asymmetrically, therefore both even- and odd-order generations happen.

To study the electron dynamics includes both inter- and intraband transitions, we need numerical solutions of the time-dependent Dirac equation (TDDE) for each occupied states in $\bm{k}$ space \cite{ikeda2020high}
\begin{equation}
\label{eq:dynamical_wave}
i\hbar\frac{\partial \psi(\bm{k},t)}{\partial t}=\hat{H}(\bm{k}+e_{l}\bm{A}(t))\psi(\bm{k},t)
\end{equation}
which omits the interactions between electrons.
The initial electron states from each band are given by the eigenvalues of the $\hat{H}$ without external field and they are evolved independently. We numerically integrate the time-dependent
Dirac equation from $-15T$ to $15T$, with applying the 4th-order
Runge-Kutta method. The outputs are confirmed almost without changes under smaller discretization.
After solving Eq.~(\ref{eq:dynamical_wave}), we can obtain the spinor wave functions $\psi_{c,v}(\bm{k},t)$ at each time and momentum point, where $c$ and $v$ stand for the conduction and valence bands, respectively.
\begin{figure}[h]
\centering
\includegraphics[width=1\linewidth]{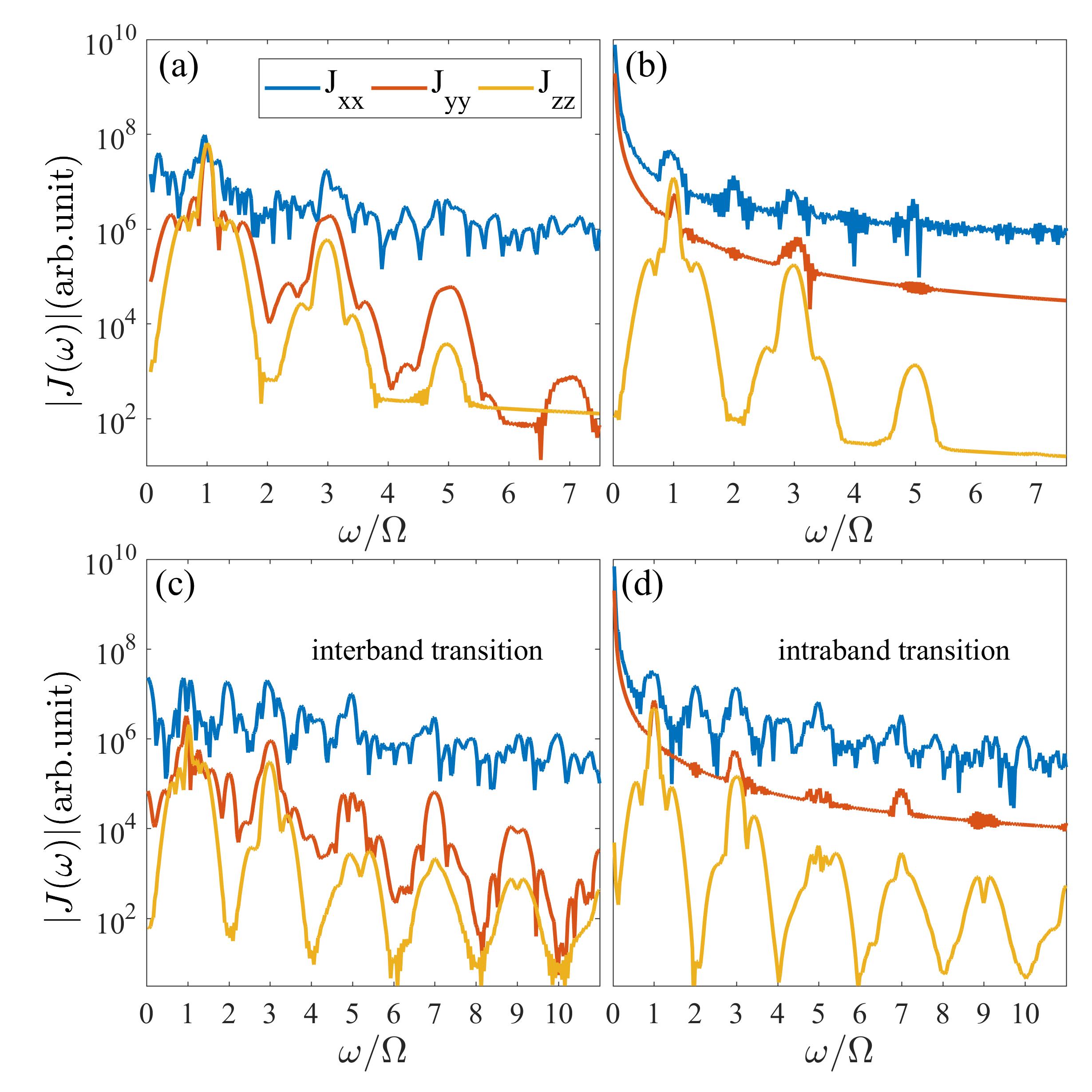}
\caption{High harmonic generations of the longitudinal currents by
inputting the electric field in three dimensions.
(a) and (b) are the spectra from the non-tilted and over-tilted Weyl cones. The strength of electric field in the simulation is $E_{0}=2.4$ kV/cm. $J_{xx}$, $J_{yy}$ and $J_{zz}$ are plotted in blue, red and yellow colors. (c) and (d) as decomposed from the HHGs of the over-tilted cones in (b), illustrate the intraband and interband transitions respectively.
}
\label{Fig2}
\end{figure}
The electric currents are defined as the expectation values of the operator $\hat{\bm{j}}=\frac{\partial\hat{H}}{\partial\bm{A}}=e\hat{\bm{v}}$,
in which group velocity matrices are defined as
$\hat{v}_{x}=\gamma A+a\hat{\sigma}_{y}+b\hat{\sigma_{z}},
\hat{v}_{y}
=\gamma B+c\hat{\sigma}_{y}+d\hat{\sigma_{z}}$ and
$\hat{v}_{z}=e\hat{\sigma}_{x}$.
In the type-II Weyl cone, the currents can be stimulated both inside
the electron pocket $(\rm{ep})$ and out of the pockets $(\rm{op})$.
In the former case, the electron states from both bands are occupied
and the currents $\bm{J}_{\rm{ep}}(t)$ can be described as
\begin{eqnarray}
\sum_{\bm{k} \in \rm{ep}} e_{l}\langle\psi_{c}(\bm{k},t)|\bm{\hat{v}}(\bm{k})|\psi_{c}(\bm{k},t)\rangle
+ e_{l}\langle\psi_{v}(\bm{k},t)|\bm{\hat{v}}(\bm{k})|\psi_{v}(\bm{k},t)\rangle.\notag
\end{eqnarray}
The currents out of pocket are only stimulated from the valence band that below the chemical potential
\begin{eqnarray}
\bm{J}_{\rm{op}}(t)=\sum_{\bm{k} \in \rm{op}} e_{l}\langle\psi_{v}(\bm{k},t)|\bm{\hat{v}}(\bm{k})|\psi_{v}(\bm{k},t)\rangle.
\end{eqnarray}
The stimulations in the whole area are $\bm{J}(t)=\bm{J}_{\rm{ep}}(t)+\bm{J}_{\rm{op}}(t)$.
\section{numerical comparison in two types of Weyl cones}
\subsection{HHG responses with  zero chemical potential}
\label{np1}
With chemical potential through the nodal energy $(E=0)$,
the non-tilted ($\gamma=0$) and over-tilted ($\gamma=1$) Weyl cones are mostly differentiated by the hyperboloidal electron and hole pocket structure. Electron states from the electron pocket are stimulated from both bands but in
noncentrosymmetrical $\bm{k}$ positions,  thus the generated $J_{\rm{ep}}(t)$ are expected to have asymmetrical responses.

Coupling with the linearly polarized pulse in each dimension, we observe the anisotropic longitudinal HHG responses as shown in Fig.~\ref{Fig2}.
In the non-tilted cone (a), the stimulated current $J_{xx}$ holds broader spectral width than other directions due to its larger Fermi velocity.
Each direction's inversion symmetry is kept in the this case, therefore we only observe the odd-order generations.
In the over-tilted case, since the states inside the electron pocket
distribute without centrosymmetry along the x and y directions, we
observe both odd- and even-order generations in these two directions. The currents perform odd-only generations in z direction where the centrosymmetry through the Weyl point is always preserved.
Significant large currents at frequency zero are generated in x and y directions, which originates from the nonlinear difference frequency process of the even-order harmonic generations.
To understand this large direct current, it is efficient to separate the HHGs as intraband and interband transitions. Here, we use the unitary transformation $\mathcal{U}(t)$ to diagonalize the instantaneous hamiltonian $\hat{H}(\bm{k},t)$.
We name the terms in Eq.~(\ref{eq:one}) as
\begin{eqnarray}
C_{0}&=&\gamma (Ak_{x}+Bk_{y});\;\ C_{1}=ek_{z}; \notag \\
C_{2}&=&ak_{x}+ck_{y};\;\;\;\;\;\;\;\;\ C_{3}=bk_{x}+dk_{y};
\end{eqnarray}
 and $L=\sqrt{C_{1}^{2}+C_{2}^{2}+C_{3}^{2}}$ for simplified purpose.
Replace ${k}_{i}$ with $k_{i}+e_{l}A_{i}(t)$ and the unitary matrix $\mathcal{U}(t)$ explicitly reads
\begin{eqnarray}
\mathcal{U}(t)=
\left(
  \begin{array}{cc}
    \frac{C_{1}(t)-iC_{2}(t)}{\sqrt{2(L(t)+C_{3}(t))L(t)}} & \frac{C_{1}(t)-iC_{2}(t)}{\sqrt{2(L(t)-C_{3}(t))L(t)}} \\
    -\frac{\sqrt{L(t)+C_{3}(t)}}{\sqrt{2L(t)}} & \frac{\sqrt{L(t)-C_{3}(t)}}{\sqrt{2L(t)}} \\
  \end{array}
\right)
\end{eqnarray}
The current operators after the same transformation are given by $\tilde{\hat{j}}_{i}=\mathcal{U}(t)^{-1}\hat{j}_{i}\mathcal{U}(t)$,
in which we denote the diagonal elements contribute to the intraband transitions and the off-diagonal elements are for the interband transitions.
Thus, the intraband and interband transitions could be decoupled.
Fig.~\ref{Fig2}(d) and (e)  illustrate the interband and intraband
transitions in the over-tilted cone, respectively.
Intuitively, the intraband transition has significant contributions to the direct currents in the over-tilted case.
This conclusion echoes the Fourier harmonics of $v_{x}(t)$ at point n, which is shown in Fig.~\ref{Fig1}(d).
The breaking of inversion symmetry is reflected at both transition processes, i.e. x- and y-direction currents show generations in all orders meanwhile the z-direction currents keep only odd generations.

\begin{figure}[ht]
\centering
\includegraphics[width=1\linewidth]{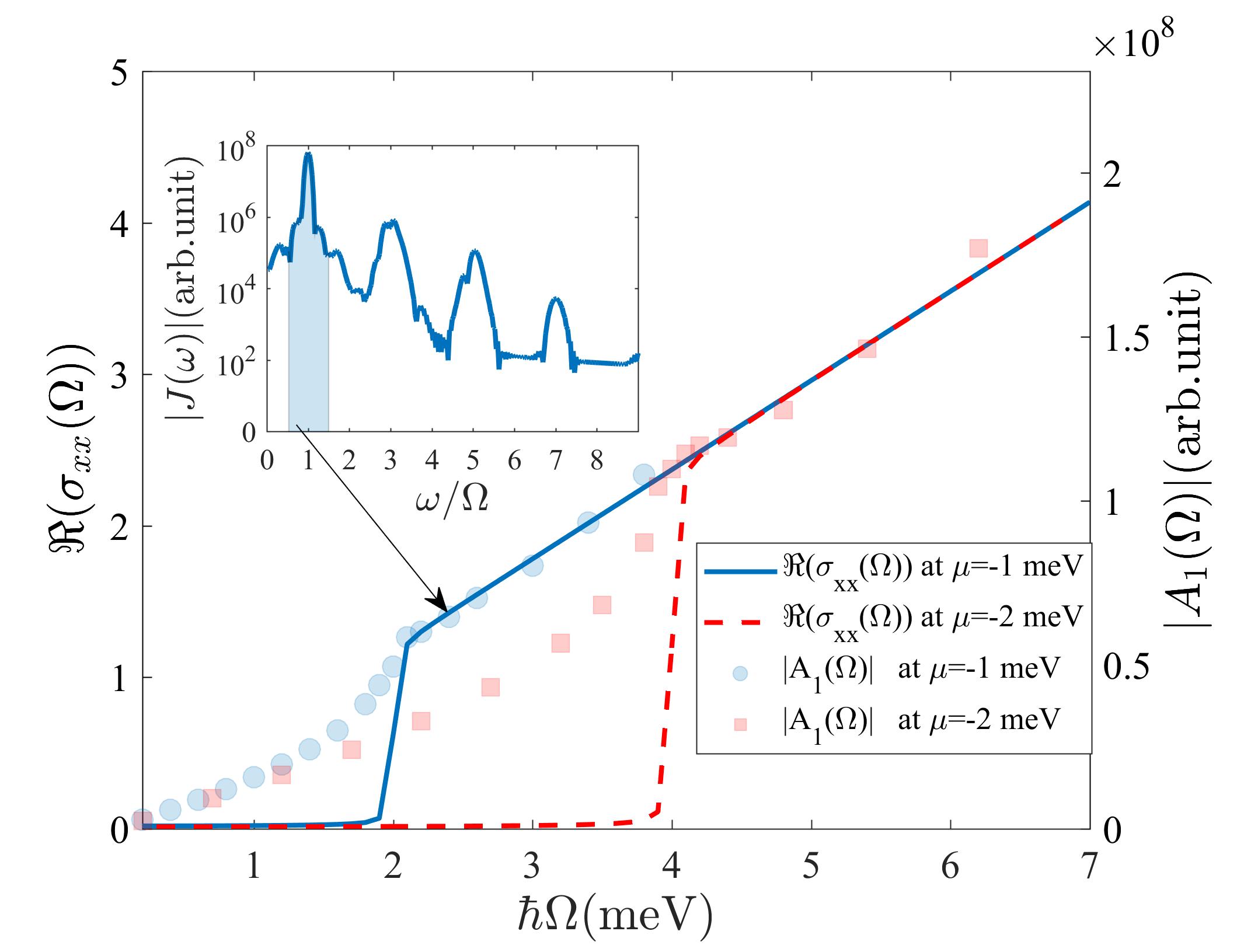}
\caption{A comparison between the linear response's strength $A_{1}(\Omega)$ and the Kubo formula's results, in a non-tilted Weyl cone.
The blue solid line and red dashed line represent the real parts of the longitudinal optical conductivity $\sigma_{xx}$ at $\mu=-1, -2$ meV. Blue circles and red squares correspond to the linear response's strength $A_{1}(\Omega)$ at $\mu=-1, -2$ meV and under different photon energy $\hbar\Omega$, in which the insert figure shows the example with $\hbar\Omega=2.4$ meV and the shadow area refers to the linear response $A_{1}(\Omega)$.
Note we take the relatively weak electric field 240V/cm to simulate the optical responses.}
\label{Fig3}
\end{figure}

It is interesting to check whether the linear response of the calculated electron currents fits the kubo formula's predication, especially under weak electric field.
This comparison is performed in a non-tilted Weyl cone by setting velocity parameters of hamiltonian in Eq.~(\ref{eq:one}) as $\gamma=0$ and $a=1;b=0;c=0;d=1;e=1$ with units $\rm{eV\AA}$.
We apply the Kubo formula as the the correlation function of velocities and spectra functions to express the  longitudinal optical conductivity in $x$ direction
\begin{eqnarray}
\sigma_{xx}(\Omega)&=&\frac{e_{l}^{2}}{\Omega\pi}\int_{-\infty}^{+\infty}
d\omega'[f(\omega')-f(\omega'+\Omega)]\notag \\ &&\times\int\frac{d\bm{k}}{(2\pi)^{3}}{\rm{Tr}}[\hat{v}_{x}\hat{A}(\bm{k},\omega')\hat{v}_{x}\hat{A}(\bm{k},\omega'+\Omega)]\notag \\
\end{eqnarray}
that $f(\omega')$ is the Fermi-Dirac function with chemical potential equals zero and temperature at $0.1$ K, $\hat{v}_{x}$ is the x-direction velocity operator and spectra function $\hat{A}(\bm{k},\omega')$ is obtained from the matrix Green function by the same derivation process in Ref.~\cite{carbotte2016}. Finally we can obtain the interband longitudinal conductivity $\sigma_{xx}$ and its real part that plotted in Figure.~\ref{Fig3} under different chemical potentials. To have a better matching with the linear response theory, we choose a weak-field pulse $240$ V/cm to simulate the currents in the x direction.
In the numerical calculations of HHGs, we define the amplitude strength $A_{n}(\Omega)=\int_{(n-1/2)\Omega}^{(n+1/2)\Omega}\frac{d\omega}{\Omega}J(\omega)$ for the n-th generation, where the linear response strength of current takes the shadow area in the insert figure.
By varying the photon energy during the stimulation,
we notice two distinct slope variations of $A_{1}$  occur exactly on $\hbar \Omega = 2$ and 4 meV positions which are twice of the chemical potentials $|\mu|$ respectively.
Comparing between the numerical results and the theoretical curves, they share the similar tendency especially when the photon energy is beyond $|2\mu|$. Such way of comparing may further expand to the exploration on the nonlinear optical responses.
\begin{figure}[h]
\centering
\includegraphics[width=1\linewidth]{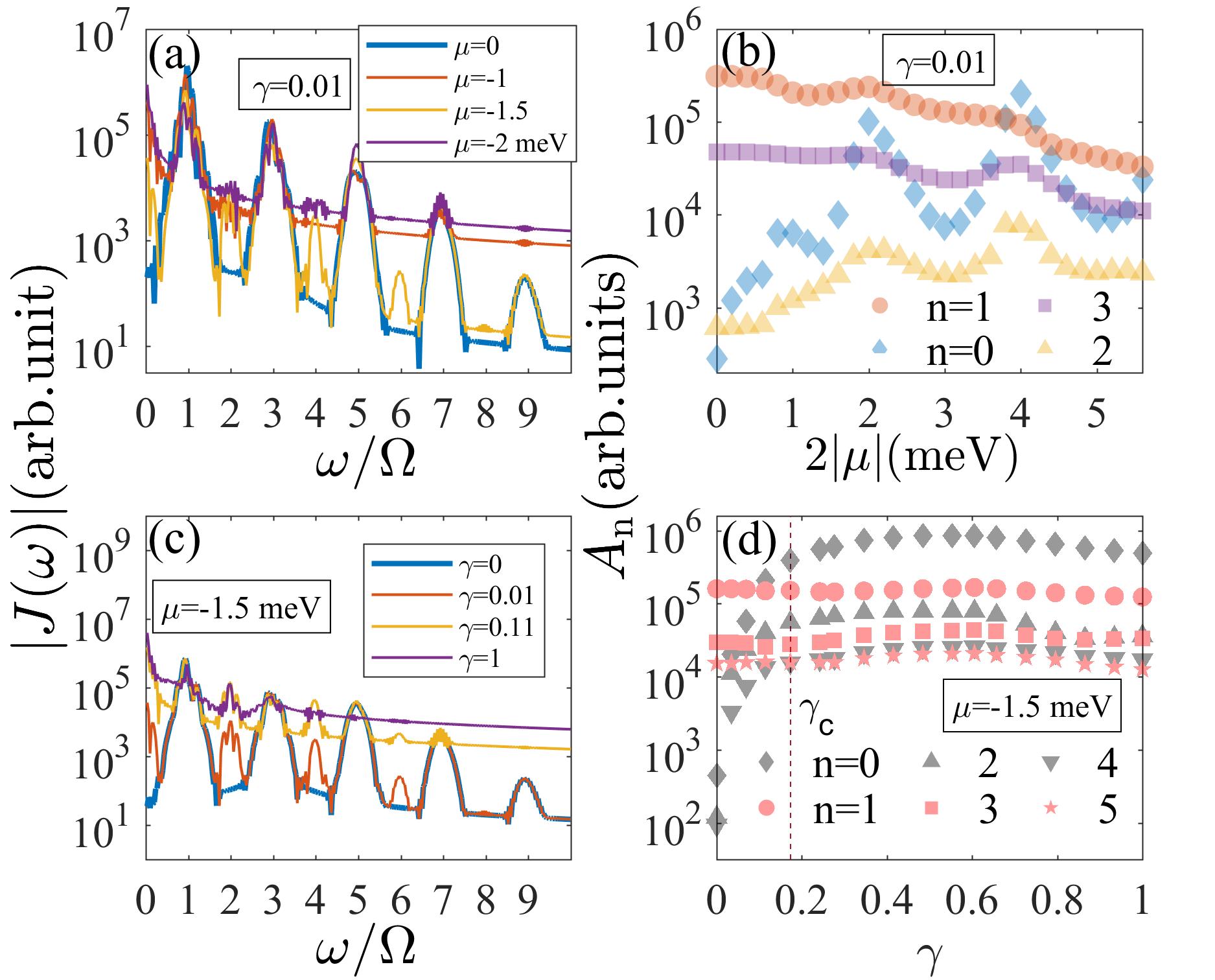}
\caption{
Subplots (a) and (b), the HHGs in a slightly tilted cone $(\gamma=0.01)$ with different chemical potentials. In (a), the curves with blue, red, yellow and purple colors refer to the chemical potential at $\mu=0, -1, -1.5$ and $-2$ meV. (b) shows how the amplitude in each order trends with the continuous change of $|2\mu|$.
(c) and (d) are the HHGs ranging from type-I to type-II cones, by the pulses with photon energy $2$ meV and chemical potential fixed at $-1.5$ meV. (c) shows the curves of Weyl cones with $\gamma=0, 0.01, 0.11, 1$ (blue, red, yellow and purple lines). In (d), the amplitude strength from each order is plotted with the changing of tilt. Dashed line marks the critical value $\gamma_{c}$ divides type-I and type-II cones.
}
\label{Fig4}
\end{figure}
\subsection{HHG responses with  non-zero chemical potential}
In the last subsection, we conclude the unusual HHG responses can reflect the asymmetric stimulations due to the hyperboloidal pocket structure in type-II cone with chemical potential at nodal energy.
To elucidate how the non-zero chemical potential influences high order generations, we evolve the electron dynamics below the nodal energy.
In this discussion, we show noncentrosymmetric
generations can be enhanced in both tilted type-I and -II cones.

Consider a slightly tilted cone $\gamma=0.01$,
and the stimulations are driven by the pulse with photon energy 2 meV and electric field strength $1.2$ kV/cm.
Figure.~\ref{Fig4}(a) shows that the even-order generations are
barely seen with zero chemical potential, which is
gradually enhanced by the chemical potential changing from $0$ to $-2$ meV. Especially, the generations in even orders are strongly
stimulated while $2|\mu|$ equals to one photon and two photons' energy. With decreasing the chemical potential continuously, we find the zero and second-order generations at $\mu=-1$ and $-2$ meV reach their highest performance, that presents evident Fermi resonances peaks of one photon and two photon process in Fig.~\ref{Fig4}(b).
When fixing the chemical potential at $-1.5$ meV,
the pick-out cases in (c) show that, with tiny tilt ($\gamma=0.01$), a remarkable enhancement of even-order generations happens. Further increasing the tilt either in type-I ($\gamma=0.11$) or type-II ($\gamma=1$), the responses approach saturation quickly.
In (d), we observe the
responses at zero, second and fourth orders rise fast
following the tilt increasing, even before the band structure is over tilted (before $\gamma_{c}$). Note the odd-order generations are not changed much with the tilt increasing.

\subsection{selective generations broken in the type-II cone}
In the limit $t_{0}\rightarrow\infty$, we can treat our
 discussing object as a time-periodic Floquet system, in which the
 dynamical symmetry is crucial \cite{neufeld2019}. As the non-perturbative Floquet process, HHGs can
 directly reflect the dynamical symmetry of the system.
\begin{figure}[h]
\centering
\includegraphics[width=1\linewidth]{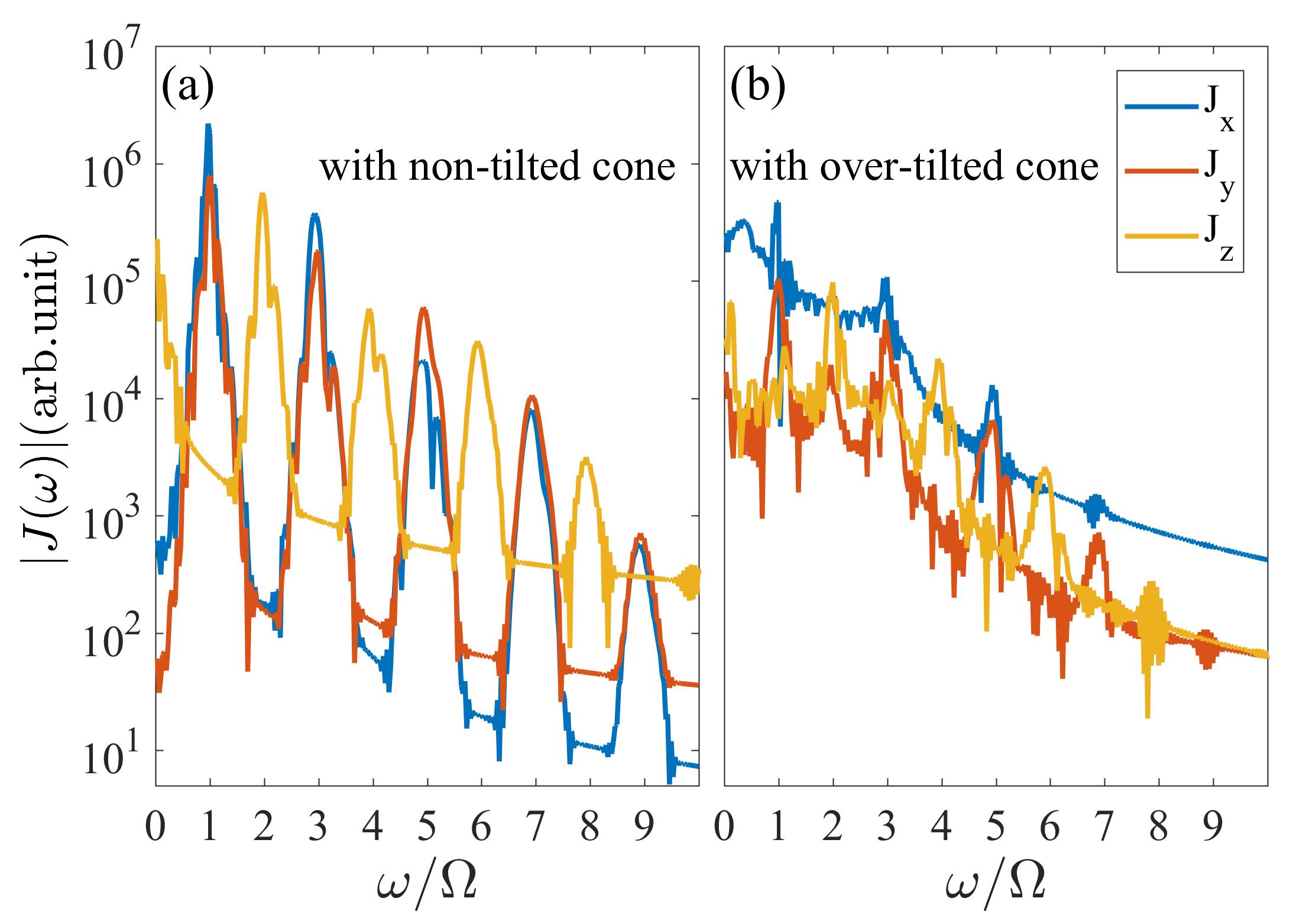}
\caption{Currents induced by the left-handed circularly polarized light in the x-y plane, with photon energy 2 meV and electric field strength 1.2 kV/cm. In the non-tilted cone (a), the odd-only generations happen in the x, y directions (blue and red lines) and even-only generations happen in the z direction (yellow line).  In the over-tilted cone (b), HHGs of the three directions are stimulated in every order.}
\label{Fig5}
\end{figure}
\begin{figure*}[ht]
\centering
\includegraphics[width=0.8\linewidth]{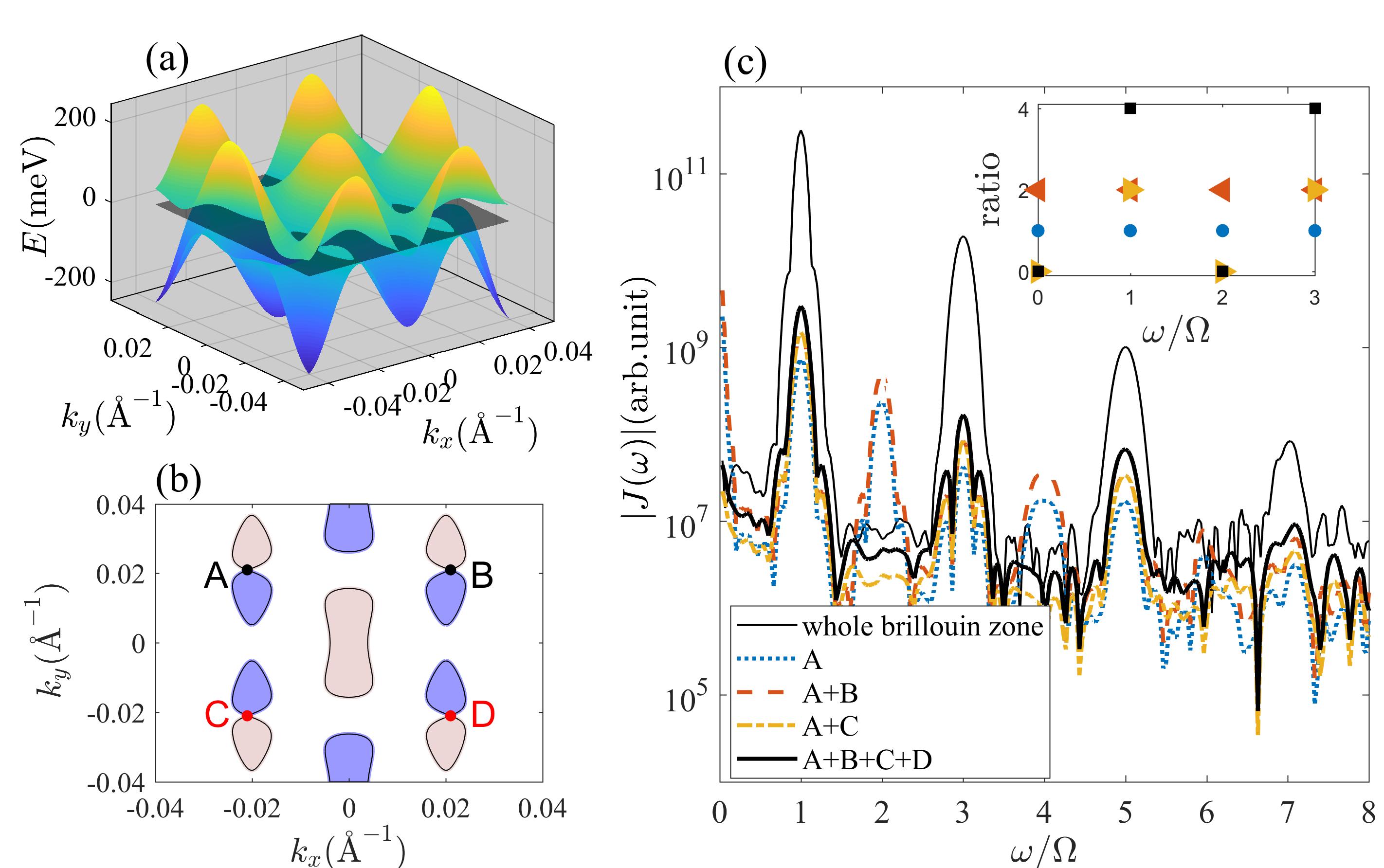}
\caption{(a) shows the band structure of a type-II Weyl semimetal, which includes two pairs of Weyl points. (b) shows the the location of electron (in blue) and hole pockets (in pink), where A and B share the same chirality that C and D have the opposed chirality.
(c) shows the y-direction currents with $\hbar\Omega=3$ meV.
In particular, dotted line shows the contribution around the electron pocket of one single Weyl point A. Dashed and dot-dashed lines show the results of two Weyl points with same chirality and with opposed chirality. The thick black line is for the total contribution of the electrons pockets around the Weyl points. The
thinner black line represents the HHGs from the whole brillouin zone. The insert figure in (c) shows the ratio of amplitude strength $A_{n}$ in different summations to $A_{n}$ containing only one single Weyl cone A (marked in circle), in which $n=0, 1, 2, 3$ refer to the harmonic orders, left triangles, right triangles, squares refer to the summations A+B, A+C and A+B+C+D with the same color indices of the lines.}
\label{Fig6}
\end{figure*}
We introduce the vector potential of a left-handed circularly polarized light into the system
\begin{equation}
\label{eq:vp2}
\bm{A}(t)=\frac{E_{0}}{\Omega}\exp[-2\ln 2(\frac{t}{{t_{\rm{0}}}})^{2}][(\vec{e}_{x}\cos(\Omega t)+ \vec{e}_{y}\sin(\Omega t)].
\end{equation}
By implanting the electric filed in x-y plane, the HHG currents at three directions are plotted in Fig.~\ref{Fig5}.
For the non-tilted system (a), the odd-only generations happen in the x, y directions and even-only generations happen in the z direction. Such phenomena of selective generations are related to the Hamiltonian symmetry $C_{2z}$, which rotates $\pi$ about the $z$ axis.  This symmetry is only preserved in hamiltonian when $\gamma=0$ in Eq.~(\ref{eq:one}).
The dynamical symmetry of the time-dependent hamiltonian needs to consider both the $C_{2z}$ symmetry and
the property of vector potential in time translations $\bm{A}(t)=-\bm{A}(t+\frac{T}{2})$
at the continuous wave approximation $(t_{0}\rightarrow\infty)$. The details of symmetry analysis to the hamiltonian, currents, states
and the selection rules' provement are shown in the Appendix \ref{dysel}.

For the over-tilted case, the hyperboloid pocket structure breaks
the $C_{2z}$ symmetry in hamiltonian and the currents can not be composed as pairwise Floquet states with opposed momentum.
With selection rules broken, we see the even- and odd-only responses in HHGs are not held anymore, as plotted in Fig.~\ref{Fig5}(b).

Dynamical symmetry has similar impact on the selective generations of transverse currents with linearly polarized light. In Figure.~\ref{transverse}(a) the non-tilted case, the x-direction transverse current $J_{xy}$ induced by the y-direction pulse show odd-only generations and the z-direction transverse current $J_{zy}$ shows even-only generations.
We notice an interesting correspondence
between the transverse currents and the
currents induced by the circularly polarized light.
Explicitly, $J_{xy}$ has the same selection rules with $J_{x}$,$J_{y}$ in Fig.~\ref{Fig5}(a) while $J_{zy}$ is consistent with the $J_{z}$ current. Such interesting correspondence
can be explained by the same dynamical analysis in the Appendix \ref{dysel}. As for the over-tilted case, $C_{2z}$ symmetry is no longer held and selective rules are broken as well.

\subsection{HHGs from the pairs of Weyl points}
In the lattice system, the 'no-go theorem' requires the Weyl fermions appear in pairs with opposite chirality.
Therefore the nonlinear responses include multiple Weyl points need to be further considered.
Here we discuss a lattice model with two pairs of node
\begin{eqnarray}
\hat{H}(\bm{k})&=&\gamma[\cos(2ak_{x})-\cos(ak_{0})][\cos(ak_{y})-\cos(ak_{0})]\hat{\sigma}_{0}\notag\\
&&-\{m[1-\cos^{2}(ak_{y})-\cos(ak_{z})]\notag\\&&+2t_{x}[\cos(ak_{x})-\cos(ak_{0})]\}\hat{\sigma}_{x}\notag\\
&&-2t\sin(ak_{z})\hat{\sigma}_{y}-2t\cos(ak_{y})\hat{\sigma}_{z}
\label{eq:weyl_pair}
\end{eqnarray}
in which four nodes located at $(\pm\pi/2a,\pm\pi/2a,0)$ break inversion symmetry \cite{mccormick2017}. In Fig.~\ref{Fig6}(a),
the over-tilted structure's parameters are chosen as $t_{x}=\frac{t}{2}$, $m=2t$ and $\gamma=2.7t$, with $t=40$ meV, $a=75 \rm{\AA}$. The pocket structures at the nodal energy are plotted in blue (electron) and pink (hole) colors In Fig.~\ref{Fig6}(b). Note the Weyl points A and B have the same chirality, while the points C and D are their counterparts with opposed chirality.

To drive the pocket electrons across the Weyl points, we applied the linearly polarized laser in y direction. Fig.~\ref{Fig6}(c) shows the local nonlinear responses around one Weyl point and their superposition effects associate with the chirality.
For a single Weyl point A, the nonlinear harmonics are strong in
each order, especially the large direct current, which agrees with the major features on the former type-II cone analysis.
Through the current amplitude ratio illustration in (c), we have a quantified understanding over the superposition effects.
While we combine the currents around two points with same chirality (A and B), the amplitude strength on each order is doubled. However, for the mixture of two points with opposed chirality (A and C), the odd-order amplitude strength are still enhanced twice, meanwhile the even-order generations are totally weakened. For example, the strength at second harmonic are two orders of magnitude weaker than the single point A, similar for the signal at frequency zero. Due the cancellation effects between opposed chirality, we can not observe strong noncentrosymmetric even-order harmonics in this type-II model as verified by the HHGs containing four points in Fig.~\ref{Fig6}(c).
Compared with the whole Brillouin zone current responses, we should note that the odd-order generations including only the Weyl points' pockets contributions show similar structures with weaker magnitude.
In the perspective of numerical calculations with discretization $100\times 100\times 100$, there are 1014766 states stimulated in whole brillouin zone, from which the 6632 states are inside the electron pockets of Weyl points. Therefore the pockets' area proportion to the whole brillouin zone has significant impact on the optical nonlinear generations.
For the non-tilted circumstance ($\gamma=0$), the Fermi surface is mainly nonsymmetric along the x direction. With pulse on this direction, we observe the stimulated even-order high harmonic generations shown in Fig.~\ref{Weyl_pair_type_I}.
\section{conclusion}
The main accomplishment of this paper is as follows.
The high harmonic generations from the WSMs are investigated both in the continuous model of a single Weyl cone and the lattice model with Weyl points in pairs.
By modelling carrier dynamics inside the hyperboloid pocket structure of type-II cone,
we observe strong noncentrosymmetric HHGs along with the large DC signal. By separating the HHGs into interband and intraband transitions, we conclude the direct currents are contributed mostly by the latter part. Meanwhile even-order generations are observed in both two transitions due to the inversion symmetry broken. In spite of tilting the cone, varying the chemical potential below nodal energy could also efficiently enhance the even-order generations which can be intuitively seen even in a slightly tilted cone. Also, strong Fermi resonances at zero- and second order harmonics happen while $|2\mu|$ changes to one or two photon energy.

The the selective generations are investigated through introducing the
left-handed circularly polarized light. The selective rule is valid in the non-tilted case since the hamiltonian's dynamical symmetry is kept. In contrast, the over-tilted structure destroys the $C_{2z}$ symmetry which induces all-order generations with no selections.

From a more realistic point of view,
we consider a lattice model that holds two pairs of Weyl points. The numerical simulations conclude the Weyl points with same chirality double the response strength, while the Weyl points with opposed chirality show enhancement in odd harmonics and cancellation in even harmonics,
i.e., the non-centrosymmetric nonlinear responses are suppressed. In the whole brillouin zone, odd-order generations dominate the optical nonlinear responses.
Our results provide a
deeper understanding of
nonlinear optical responses based on the remarkable features of
tilted semimetals and we will promote the study to other novel structures such as nodal line semimetal, hybrid semimetal, etc.

\begin{acknowledgments}
We thank Tatsuhiko N. Ikeda for
illuminating discussions.
The work is performed at the Chinese Academy of Science Terahertz Science Center, which is supported by National Natural Science Foundation of China Grant No. 61988102.
\end{acknowledgments}

\appendix
\renewcommand\thefigure{\Alph{section}\arabic{figure}}

\setcounter{figure}{0}

\section{Dynamical symmetry and selection rules}
\label{dysel}
The non-tilted cone's hamiltonian ($\gamma=0$) in Eq.~(\ref{eq:one}) preserves symmetry $C_{2z}$, i.e.
\begin{eqnarray}
h(\bm{k})=C_{2z}h(\bm{k}')C_{2z}^{\dag}
\end{eqnarray}
where $\bm{k}'=(-k_{x},-k_{y},k_{z})$.
In the following discussions, we use the approximation $t_{0}\rightarrow \infty$ that $\bm{A}(t)$ holds continuous wave's form and have the time translation property $\bm{A}(t)=-\bm{A}(t+\frac{T}{2})$.

Consider a circularly polarized light in x-y plane is implanted into the system and the time-dependent hamiltonian $h(\bm{k},t)$ preserves such dynamical symmetry
\begin{eqnarray}\label{eq:dy_th}
h(\bm{k},t)&=C_{2z}h(\bm{k}',t+T/2)C_{2z}^{\dag}.
\end{eqnarray}
The currents under such operation satisfy
\begin{eqnarray}
j_{x}(\bm{k},t)&&=-C_{2z}j_{x}(\bm{k}',t+T/2)C_{2z}^{\dag};\notag \\
j_{y}(\bm{k},t)&&=-C_{2z}j_{y}(\bm{k}',t+T/2)C_{2z}^{\dag};\notag \\
j_{z}(\bm{k},t)&&=C_{2z}j_{z}(\bm{k}',t+T/2)C_{2z}^{\dag}.\label{eq:dy_hc}
\end{eqnarray}

The time-dependent quantum equation (TDQE) is
\begin{eqnarray}\label{eq:td_q}
i\hbar\frac{\partial \vec{\psi}(\bm{k},t)}{\partial t}=h(\bm{k},t)\vec{\psi}(\bm{k},t).
\end{eqnarray}
Substitute the dynamical symmetry relations Eq.~(\ref{eq:dy_th}) into Eq.~(\ref{eq:td_q}) and  do the time translation $t\rightarrow t-\frac{T}{2}$, then it changes to be
\begin{eqnarray}
i\hbar\frac{\partial C_{2z}^{\dag}\vec{\psi}(\bm{k},t-\frac{T}{2})}{\partial t}=h(\bm{k}',t)C_{2z}^{\dag}\vec{\psi}(\bm{k},t-\frac{T}{2}).
\label{eq:dynamical}
\end{eqnarray}
The TDQE at $\bm{k}'$ point is written as
\begin{eqnarray}\label{eq:td_qk}
i\hbar\frac{\partial \vec{\psi}(\bm{k}',t)}{\partial t}=h(\bm{k}',t)\vec{\psi}(\bm{k}',t).
\end{eqnarray}
Comparing Eq.~(\ref{eq:dynamical}) with Eq.~(\ref{eq:td_qk}) and suppose there is no degeneracy, then the eigenfunctions are connected as
\begin{eqnarray}\label{eq:td_w}
\vec{\psi}(\bm{k}',t)=C_{2z}^{\dag}\vec{\psi}(\bm{k},t-\frac{T}{2}).
\end{eqnarray}
Owing to the periodicity in time, the general solution of TDQE in Floquet theorem is
\begin{eqnarray}
\vec{\psi}^{F}_{m}(\bm{k},t)=e^{-iE_{m}^{F}(\bm{k})t}\vec{u}_{m}(\bm{k},t)
\end{eqnarray}
where $E_{m}^{F}(\bm{k})$ is the Floquet eigenvalues that
satisfies $E_{m}^{F}(\bm{k})=E_{m}^{F}(\bm{k}')$ in the non-tilted case, $\vec{u}_{m}(\bm{k},t)$ is a T-periodical function and $m$ is the band index. Through the derivations Eq.~(\ref{eq:td_q})-(\ref{eq:td_qk}), we obtain the relations of Floquet states just like Eq.~(\ref{eq:td_w})
\begin{eqnarray}\label{eq:td_F}
\vec{\psi}_{m}^{F}(\bm{k}',t)=C_{2z}^{\dag}\vec{\psi}_{m}^{F}(\bm{k},t-\frac{T}{2}).
\end{eqnarray}
\begin{figure}[h]
\centering
\includegraphics[width=0.9\linewidth]{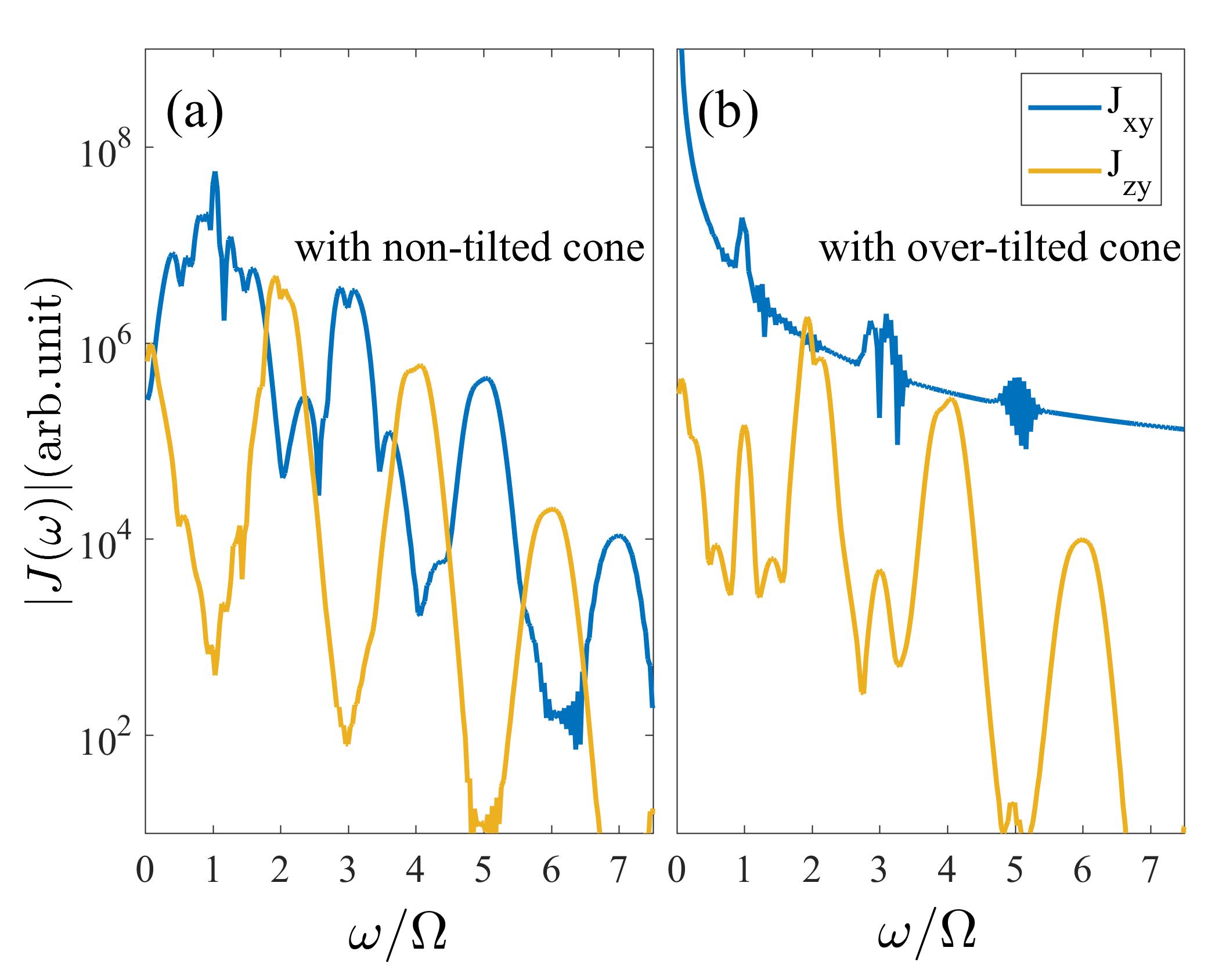}
\caption{Transverse currents stimulated by the y-direction polarized pulse in the non-tilted ($\gamma=0$) and over-tilted ($\gamma=1$) cones.
In subplot (a), the transverse current $J_{xy}$ in x direction shows odd-only generations (blue line) and
current $J_{zy}$ in z direction shows even-only generations (yellow line) .
The correspondent cases in the over-tilted cone ($\gamma=1$) are shown in (b).}
\label{transverse}
\end{figure}
We project the electric currents $j_{\alpha}$ into the $m$-th pairwise Floquet states
\begin{eqnarray}\label{eq:c}
&&J_{m}^{\alpha}(\bm{k},t)\notag\\&=&\vec{\psi}^{F\dag}_{m}(\bm{k},t)j_{\alpha}(\bm{k},t)\vec{\psi}^{F}_{m}(\bm{k},t)
+\vec{\psi}^{F\dag}_{m}(\bm{k}',t)j_{\alpha}(\bm{k}',t)\vec{\psi}^{F}_{m}(\bm{k}',t). \notag\\
\end{eqnarray}
Substitute Eq.~(\ref{eq:dy_hc}) and Eq.~(\ref{eq:td_F}) into Eq.~(\ref{eq:c}), we find the connection as follows
\begin{align}
J_{m}^{x}(\bm{k},t)&=-J_{m}^{x}(\bm{k},t+T/2);\notag\\
J_{m}^{y}(\bm{k},t)&=-J_{m}^{y}(\bm{k},t+T/2);\notag\\
J_{m}^{z}(\bm{k},t)&=J_{m}^{z}(\bm{k},t+T/2).
\label{eq:Jt}
\end{align}
The n-th order harmonic generation in x direction with considering the symmetry of currents in Eq.~(\ref{eq:Jt}) is
\begin{eqnarray}
J_{m}^{x}(\bm{k},n\Omega)&=&\int_{0}^{T}\frac{dt}{T}e^{in\Omega t}J_{m}^{x}(\bm{k},t)\notag\\
&=&\int_{0}^{T}\frac{dt}{T}e^{in\Omega (t+\frac{T}{2})}J_{m}^{x}(\bm{k},t+\frac{T}{2})\notag\\
&=&-e^{in\pi}J_{m}^{x}(\bm{k},n\Omega).
\end{eqnarray}
Similarly we obtain $J_{m}^{y}(\bm{k},n\Omega)=-e^{in\pi}J_{m}^{y}(\bm{k},n\Omega)$ and $J_{m}^{z}(\bm{k},n\Omega)=e^{in\pi}J_{m}^{z}(\bm{k},n\Omega)$.
The selection rules are obvious: $J_{m}^{x,y}(\bm{k},n\Omega)=0$ when $n$ is even number;
$J_{m}^{z}(\bm{k},n\Omega)=0$ when $n$ is odd number. Sum the currents in each band and we will have the same selective generations.
In the type-II cones, the symmetry $C_{2z}$ is not held anymore and we see the HHGs happen in each order.
The former dynamical analysis about circularly polarized light can also be applied into the linearly polarized case. In the non-tilted case, the transverse currents stimulated by the y-direction polarized light also show selective generations that odd orders in the x direction and even orders in the z direction (Fig.~\ref{transverse}). In the over-tilted generations, we see in Fig.~\ref{transverse}(b) that selective generations are destroyed without holding the dynamical symmetry.
\section{nonlinear optical responses in a non-tilted lattice model}
\begin{figure}[h]
\centering
\includegraphics[width=0.9\linewidth]{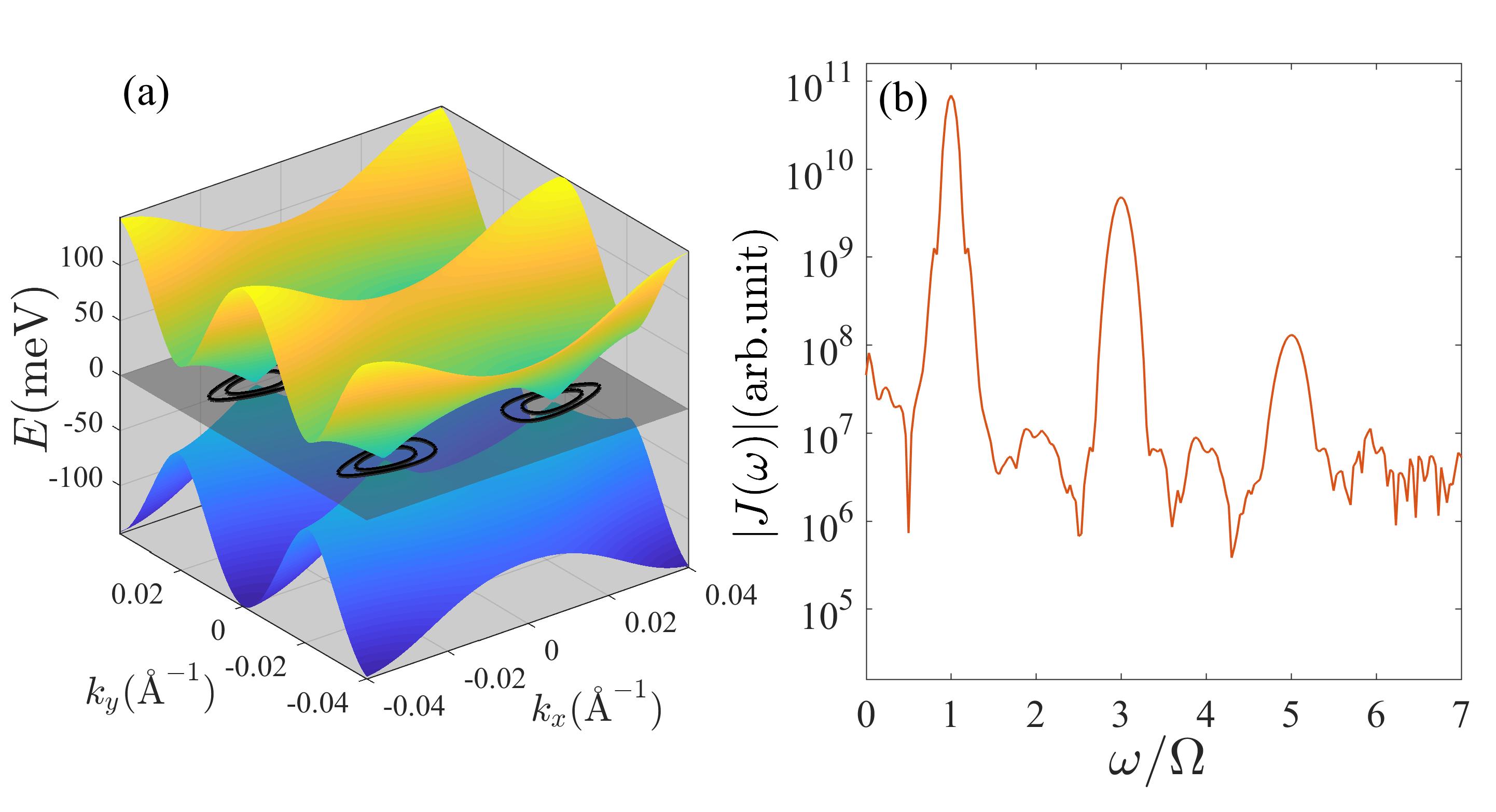}
\caption{(a) shows the band structure of the non-tilted type-I Weyl semimetal at $k_{z}=0$ plane, described in Eq.~(\ref{eq:weyl_pair}). The contour lines show the distorted Fermi surfaces around the Weyl points that asymmetric along x direction. (b) The HHGs are stimulated by the x-polarized pulse with energy $\hbar\Omega=3$ meV. Odd- and even-order generations are all stimulated, with the broken inversion symmetry.}
\label{Weyl_pair_type_I}
\end{figure}

Figure.~\ref{Weyl_pair_type_I} shows the nonlinear optical responses in the non-tilted lattice model (with $\gamma=0$ in Eq.~(\ref{eq:weyl_pair})). The Fermi surface around each Weyl point is distorted and asymmetric along the x direction. The longitudinal HHGs illustrate the even-order responses in this direction besides the intense odd-order generations.

\nocite{*}

\bibliography{apssamp}

\begin{thebibliography}{65}
\expandafter\ifx\csname natexlab\endcsname\relax\def\natexlab#1{#1}\fi
\expandafter\ifx\csname bibnamefont\endcsname\relax
  \def\bibnamefont#1{#1}\fi
\expandafter\ifx\csname bibfnamefont\endcsname\relax
  \def\bibfnamefont#1{#1}\fi
\expandafter\ifx\csname citenamefont\endcsname\relax
  \def\citenamefont#1{#1}\fi
\expandafter\ifx\csname url\endcsname\relax
  \def\url#1{\texttt{#1}}\fi
\expandafter\ifx\csname urlprefix\endcsname\relax\def\urlprefix{URL }\fi
\providecommand{\bibinfo}[2]{#2}
\providecommand{\eprint}[2][]{\url{#2}}

\bibitem[{\citenamefont{Boyd}(2020)}]{boyd2020}
\bibinfo{author}{\bibfnamefont{R.~W.} \bibnamefont{Boyd}},
  \emph{\bibinfo{title}{Nonlinear optics}} (\bibinfo{publisher}{Academic
  press}, \bibinfo{year}{2020}).

\bibitem[{\citenamefont{McPherson et~al.}(1987)\citenamefont{McPherson, Gibson,
  Jara, Johann, Luk, McIntyre, Boyer, and Rhodes}}]{mcpherson1987}
\bibinfo{author}{\bibfnamefont{A.}~\bibnamefont{McPherson}},
  \bibinfo{author}{\bibfnamefont{G.}~\bibnamefont{Gibson}},
  \bibinfo{author}{\bibfnamefont{H.}~\bibnamefont{Jara}},
  \bibinfo{author}{\bibfnamefont{U.}~\bibnamefont{Johann}},
  \bibinfo{author}{\bibfnamefont{T.~S.} \bibnamefont{Luk}},
  \bibinfo{author}{\bibfnamefont{I.}~\bibnamefont{McIntyre}},
  \bibinfo{author}{\bibfnamefont{K.}~\bibnamefont{Boyer}}, \bibnamefont{and}
  \bibinfo{author}{\bibfnamefont{C.~K.} \bibnamefont{Rhodes}},
  \bibinfo{journal}{Journal of the Optical Society of America B}
  \textbf{\bibinfo{volume}{4}}, \bibinfo{pages}{595} (\bibinfo{year}{1987}).

\bibitem[{\citenamefont{Krause et~al.}(1992)\citenamefont{Krause, Schafer, and
  Kulander}}]{krause1992}
\bibinfo{author}{\bibfnamefont{J.~L.} \bibnamefont{Krause}},
  \bibinfo{author}{\bibfnamefont{K.~J.} \bibnamefont{Schafer}},
  \bibnamefont{and} \bibinfo{author}{\bibfnamefont{K.~C.}
  \bibnamefont{Kulander}}, \bibinfo{journal}{Physical Review Letters}
  \textbf{\bibinfo{volume}{68}}, \bibinfo{pages}{3535} (\bibinfo{year}{1992}).

\bibitem[{\citenamefont{Schafer et~al.}(1993)\citenamefont{Schafer, Yang,
  DiMauro, and Kulander}}]{schafer1993}
\bibinfo{author}{\bibfnamefont{K.}~\bibnamefont{Schafer}},
  \bibinfo{author}{\bibfnamefont{B.}~\bibnamefont{Yang}},
  \bibinfo{author}{\bibfnamefont{L.}~\bibnamefont{DiMauro}}, \bibnamefont{and}
  \bibinfo{author}{\bibfnamefont{K.}~\bibnamefont{Kulander}},
  \bibinfo{journal}{Physical Review Letters} \textbf{\bibinfo{volume}{70}},
  \bibinfo{pages}{1599} (\bibinfo{year}{1993}).

\bibitem[{\citenamefont{Corkum}(1993)}]{corkum1993}
\bibinfo{author}{\bibfnamefont{P.~B.} \bibnamefont{Corkum}},
  \bibinfo{journal}{Physical Review Letters} \textbf{\bibinfo{volume}{71}},
  \bibinfo{pages}{1994} (\bibinfo{year}{1993}).

\bibitem[{\citenamefont{Ghimire and Reis}(2019)}]{ghimire2019}
\bibinfo{author}{\bibfnamefont{S.}~\bibnamefont{Ghimire}} \bibnamefont{and}
  \bibinfo{author}{\bibfnamefont{D.~A.} \bibnamefont{Reis}},
  \bibinfo{journal}{Nature physics} \textbf{\bibinfo{volume}{15}},
  \bibinfo{pages}{10} (\bibinfo{year}{2019}).

\bibitem[{\citenamefont{Ghimire et~al.}(2011)\citenamefont{Ghimire, DiChiara,
  Sistrunk, Agostini, DiMauro, and Reis}}]{ghimire2011}
\bibinfo{author}{\bibfnamefont{S.}~\bibnamefont{Ghimire}},
  \bibinfo{author}{\bibfnamefont{A.~D.} \bibnamefont{DiChiara}},
  \bibinfo{author}{\bibfnamefont{E.}~\bibnamefont{Sistrunk}},
  \bibinfo{author}{\bibfnamefont{P.}~\bibnamefont{Agostini}},
  \bibinfo{author}{\bibfnamefont{L.~F.} \bibnamefont{DiMauro}},
  \bibnamefont{and} \bibinfo{author}{\bibfnamefont{D.~A.} \bibnamefont{Reis}},
  \bibinfo{journal}{Nature physics} \textbf{\bibinfo{volume}{7}},
  \bibinfo{pages}{138} (\bibinfo{year}{2011}).

\bibitem[{\citenamefont{Vampa and Brabec}(2017)}]{vampa2017}
\bibinfo{author}{\bibfnamefont{G.}~\bibnamefont{Vampa}} \bibnamefont{and}
  \bibinfo{author}{\bibfnamefont{T.}~\bibnamefont{Brabec}},
  \bibinfo{journal}{Journal of Physics B: Atomic, Molecular and Optical
  Physics} \textbf{\bibinfo{volume}{50}}, \bibinfo{pages}{083001}
  (\bibinfo{year}{2017}).

\bibitem[{\citenamefont{Garg et~al.}(2018)\citenamefont{Garg, Kim, and
  Goulielmakis}}]{garg2018}
\bibinfo{author}{\bibfnamefont{M.}~\bibnamefont{Garg}},
  \bibinfo{author}{\bibfnamefont{H.-Y.} \bibnamefont{Kim}}, \bibnamefont{and}
  \bibinfo{author}{\bibfnamefont{E.}~\bibnamefont{Goulielmakis}},
  \bibinfo{journal}{Nature Photonics} \textbf{\bibinfo{volume}{12}},
  \bibinfo{pages}{291} (\bibinfo{year}{2018}).

\bibitem[{\citenamefont{Li et~al.}(2020{\natexlab{a}})\citenamefont{Li, Lu,
  Chew, Han, Li, Wu, Wang, Ghimire, and Chang}}]{li2020}
\bibinfo{author}{\bibfnamefont{J.}~\bibnamefont{Li}},
  \bibinfo{author}{\bibfnamefont{J.}~\bibnamefont{Lu}},
  \bibinfo{author}{\bibfnamefont{A.}~\bibnamefont{Chew}},
  \bibinfo{author}{\bibfnamefont{S.}~\bibnamefont{Han}},
  \bibinfo{author}{\bibfnamefont{J.}~\bibnamefont{Li}},
  \bibinfo{author}{\bibfnamefont{Y.}~\bibnamefont{Wu}},
  \bibinfo{author}{\bibfnamefont{H.}~\bibnamefont{Wang}},
  \bibinfo{author}{\bibfnamefont{S.}~\bibnamefont{Ghimire}}, \bibnamefont{and}
  \bibinfo{author}{\bibfnamefont{Z.}~\bibnamefont{Chang}},
  \bibinfo{journal}{Nature Communications} \textbf{\bibinfo{volume}{11}},
  \bibinfo{pages}{1} (\bibinfo{year}{2020}{\natexlab{a}}).

\bibitem[{\citenamefont{Sivis et~al.}(2013)\citenamefont{Sivis, Duwe, Abel, and
  Ropers}}]{sivis2013}
\bibinfo{author}{\bibfnamefont{M.}~\bibnamefont{Sivis}},
  \bibinfo{author}{\bibfnamefont{M.}~\bibnamefont{Duwe}},
  \bibinfo{author}{\bibfnamefont{B.}~\bibnamefont{Abel}}, \bibnamefont{and}
  \bibinfo{author}{\bibfnamefont{C.}~\bibnamefont{Ropers}},
  \bibinfo{journal}{Nature Physics} \textbf{\bibinfo{volume}{9}},
  \bibinfo{pages}{304} (\bibinfo{year}{2013}).

\bibitem[{\citenamefont{Han et~al.}(2016)\citenamefont{Han, Kim, Kim, Kim, Kim,
  Park, and Kim}}]{han2016}
\bibinfo{author}{\bibfnamefont{S.}~\bibnamefont{Han}},
  \bibinfo{author}{\bibfnamefont{H.}~\bibnamefont{Kim}},
  \bibinfo{author}{\bibfnamefont{Y.~W.} \bibnamefont{Kim}},
  \bibinfo{author}{\bibfnamefont{Y.-J.} \bibnamefont{Kim}},
  \bibinfo{author}{\bibfnamefont{S.}~\bibnamefont{Kim}},
  \bibinfo{author}{\bibfnamefont{I.-Y.} \bibnamefont{Park}}, \bibnamefont{and}
  \bibinfo{author}{\bibfnamefont{S.-W.} \bibnamefont{Kim}},
  \bibinfo{journal}{Nature Communications} \textbf{\bibinfo{volume}{7}},
  \bibinfo{pages}{1} (\bibinfo{year}{2016}).

\bibitem[{\citenamefont{Vampa et~al.}(2017)\citenamefont{Vampa, Ghamsari,
  Siadat~Mousavi, Hammond, Olivieri, Lisicka-Skrek, Naumov, Villeneuve,
  Staudte, Berini et~al.}}]{vampa2017plasmon}
\bibinfo{author}{\bibfnamefont{G.}~\bibnamefont{Vampa}},
  \bibinfo{author}{\bibfnamefont{B.}~\bibnamefont{Ghamsari}},
  \bibinfo{author}{\bibfnamefont{S.}~\bibnamefont{Siadat~Mousavi}},
  \bibinfo{author}{\bibfnamefont{T.}~\bibnamefont{Hammond}},
  \bibinfo{author}{\bibfnamefont{A.}~\bibnamefont{Olivieri}},
  \bibinfo{author}{\bibfnamefont{E.}~\bibnamefont{Lisicka-Skrek}},
  \bibinfo{author}{\bibfnamefont{A.~Y.} \bibnamefont{Naumov}},
  \bibinfo{author}{\bibfnamefont{D.}~\bibnamefont{Villeneuve}},
  \bibinfo{author}{\bibfnamefont{A.}~\bibnamefont{Staudte}},
  \bibinfo{author}{\bibfnamefont{P.}~\bibnamefont{Berini}},
  \bibnamefont{et~al.}, \bibinfo{journal}{Nature Physics}
  \textbf{\bibinfo{volume}{13}}, \bibinfo{pages}{659} (\bibinfo{year}{2017}).

\bibitem[{\citenamefont{Vampa et~al.}(2015{\natexlab{a}})\citenamefont{Vampa,
  Hammond, Thir{\'e}, Schmidt, L{\'e}gar{\'e}, McDonald, Brabec, Klug, and
  Corkum}}]{vampa2015all}
\bibinfo{author}{\bibfnamefont{G.}~\bibnamefont{Vampa}},
  \bibinfo{author}{\bibfnamefont{T.}~\bibnamefont{Hammond}},
  \bibinfo{author}{\bibfnamefont{N.}~\bibnamefont{Thir{\'e}}},
  \bibinfo{author}{\bibfnamefont{B.}~\bibnamefont{Schmidt}},
  \bibinfo{author}{\bibfnamefont{F.}~\bibnamefont{L{\'e}gar{\'e}}},
  \bibinfo{author}{\bibfnamefont{C.}~\bibnamefont{McDonald}},
  \bibinfo{author}{\bibfnamefont{T.}~\bibnamefont{Brabec}},
  \bibinfo{author}{\bibfnamefont{D.}~\bibnamefont{Klug}}, \bibnamefont{and}
  \bibinfo{author}{\bibfnamefont{P.}~\bibnamefont{Corkum}},
  \bibinfo{journal}{Physical Review Letters} \textbf{\bibinfo{volume}{115}},
  \bibinfo{pages}{193603} (\bibinfo{year}{2015}{\natexlab{a}}).

\bibitem[{\citenamefont{Tancogne-Dejean
  et~al.}(2017{\natexlab{a}})\citenamefont{Tancogne-Dejean, M{\"u}cke,
  K{\"a}rtner, and Rubio}}]{tancogne2017}
\bibinfo{author}{\bibfnamefont{N.}~\bibnamefont{Tancogne-Dejean}},
  \bibinfo{author}{\bibfnamefont{O.~D.} \bibnamefont{M{\"u}cke}},
  \bibinfo{author}{\bibfnamefont{F.~X.} \bibnamefont{K{\"a}rtner}},
  \bibnamefont{and} \bibinfo{author}{\bibfnamefont{A.}~\bibnamefont{Rubio}},
  \bibinfo{journal}{Physical Review Letters} \textbf{\bibinfo{volume}{118}},
  \bibinfo{pages}{087403} (\bibinfo{year}{2017}{\natexlab{a}}).

\bibitem[{\citenamefont{Lanin et~al.}(2017)\citenamefont{Lanin, Stepanov,
  Fedotov, and Zheltikov}}]{lanin2017}
\bibinfo{author}{\bibfnamefont{A.}~\bibnamefont{Lanin}},
  \bibinfo{author}{\bibfnamefont{E.}~\bibnamefont{Stepanov}},
  \bibinfo{author}{\bibfnamefont{A.}~\bibnamefont{Fedotov}}, \bibnamefont{and}
  \bibinfo{author}{\bibfnamefont{A.}~\bibnamefont{Zheltikov}},
  \bibinfo{journal}{Optica} \textbf{\bibinfo{volume}{4}}, \bibinfo{pages}{516}
  (\bibinfo{year}{2017}).

\bibitem[{\citenamefont{Li et~al.}(2020{\natexlab{b}})\citenamefont{Li, Lan,
  He, Cao, Zhang, and Lu}}]{li2020determination}
\bibinfo{author}{\bibfnamefont{L.}~\bibnamefont{Li}},
  \bibinfo{author}{\bibfnamefont{P.}~\bibnamefont{Lan}},
  \bibinfo{author}{\bibfnamefont{L.}~\bibnamefont{He}},
  \bibinfo{author}{\bibfnamefont{W.}~\bibnamefont{Cao}},
  \bibinfo{author}{\bibfnamefont{Q.}~\bibnamefont{Zhang}}, \bibnamefont{and}
  \bibinfo{author}{\bibfnamefont{P.}~\bibnamefont{Lu}},
  \bibinfo{journal}{Physical Review Letters} \textbf{\bibinfo{volume}{124}},
  \bibinfo{pages}{157403} (\bibinfo{year}{2020}{\natexlab{b}}).

\bibitem[{\citenamefont{Silva et~al.}(2019)\citenamefont{Silva,
  Jim{\'e}nez-Gal{\'a}n, Amorim, Smirnova, and Ivanov}}]{silva2019}
\bibinfo{author}{\bibfnamefont{R.}~\bibnamefont{Silva}},
  \bibinfo{author}{\bibfnamefont{{\'A}.}~\bibnamefont{Jim{\'e}nez-Gal{\'a}n}},
  \bibinfo{author}{\bibfnamefont{B.}~\bibnamefont{Amorim}},
  \bibinfo{author}{\bibfnamefont{O.}~\bibnamefont{Smirnova}}, \bibnamefont{and}
  \bibinfo{author}{\bibfnamefont{M.}~\bibnamefont{Ivanov}},
  \bibinfo{journal}{Nature Photonics} \textbf{\bibinfo{volume}{13}},
  \bibinfo{pages}{849} (\bibinfo{year}{2019}).

\bibitem[{\citenamefont{Chac{\'o}n et~al.}(2020)\citenamefont{Chac{\'o}n, Kim,
  Zhu, Kelly, Dauphin, Pisanty, Maxwell, Pic{\'o}n, Ciappina, Kim
  et~al.}}]{chacon2020}
\bibinfo{author}{\bibfnamefont{A.}~\bibnamefont{Chac{\'o}n}},
  \bibinfo{author}{\bibfnamefont{D.}~\bibnamefont{Kim}},
  \bibinfo{author}{\bibfnamefont{W.}~\bibnamefont{Zhu}},
  \bibinfo{author}{\bibfnamefont{S.~P.} \bibnamefont{Kelly}},
  \bibinfo{author}{\bibfnamefont{A.}~\bibnamefont{Dauphin}},
  \bibinfo{author}{\bibfnamefont{E.}~\bibnamefont{Pisanty}},
  \bibinfo{author}{\bibfnamefont{A.~S.} \bibnamefont{Maxwell}},
  \bibinfo{author}{\bibfnamefont{A.}~\bibnamefont{Pic{\'o}n}},
  \bibinfo{author}{\bibfnamefont{M.~F.} \bibnamefont{Ciappina}},
  \bibinfo{author}{\bibfnamefont{D.~E.} \bibnamefont{Kim}},
  \bibnamefont{et~al.}, \bibinfo{journal}{Physical Review B}
  \textbf{\bibinfo{volume}{102}}, \bibinfo{pages}{134115}
  (\bibinfo{year}{2020}).

\bibitem[{\citenamefont{Schmid et~al.}(2021)\citenamefont{Schmid, Weigl,
  Gr{\"o}ssing, Junk, Gorini, Schlauderer, Ito, Meierhofer, Hofmann, Afanasiev
  et~al.}}]{schmid2021}
\bibinfo{author}{\bibfnamefont{C.~P.} \bibnamefont{Schmid}},
  \bibinfo{author}{\bibfnamefont{L.}~\bibnamefont{Weigl}},
  \bibinfo{author}{\bibfnamefont{P.}~\bibnamefont{Gr{\"o}ssing}},
  \bibinfo{author}{\bibfnamefont{V.}~\bibnamefont{Junk}},
  \bibinfo{author}{\bibfnamefont{C.}~\bibnamefont{Gorini}},
  \bibinfo{author}{\bibfnamefont{S.}~\bibnamefont{Schlauderer}},
  \bibinfo{author}{\bibfnamefont{S.}~\bibnamefont{Ito}},
  \bibinfo{author}{\bibfnamefont{M.}~\bibnamefont{Meierhofer}},
  \bibinfo{author}{\bibfnamefont{N.}~\bibnamefont{Hofmann}},
  \bibinfo{author}{\bibfnamefont{D.}~\bibnamefont{Afanasiev}},
  \bibnamefont{et~al.}, \bibinfo{journal}{Nature}
  \textbf{\bibinfo{volume}{593}}, \bibinfo{pages}{385} (\bibinfo{year}{2021}).

\bibitem[{\citenamefont{Bai et~al.}(2021)\citenamefont{Bai, Fei, Wang, Li, Li,
  Song, Li, Xu, and Liu}}]{bai2021}
\bibinfo{author}{\bibfnamefont{Y.}~\bibnamefont{Bai}},
  \bibinfo{author}{\bibfnamefont{F.}~\bibnamefont{Fei}},
  \bibinfo{author}{\bibfnamefont{S.}~\bibnamefont{Wang}},
  \bibinfo{author}{\bibfnamefont{N.}~\bibnamefont{Li}},
  \bibinfo{author}{\bibfnamefont{X.}~\bibnamefont{Li}},
  \bibinfo{author}{\bibfnamefont{F.}~\bibnamefont{Song}},
  \bibinfo{author}{\bibfnamefont{R.}~\bibnamefont{Li}},
  \bibinfo{author}{\bibfnamefont{Z.}~\bibnamefont{Xu}}, \bibnamefont{and}
  \bibinfo{author}{\bibfnamefont{P.}~\bibnamefont{Liu}},
  \bibinfo{journal}{Nature Physics} \textbf{\bibinfo{volume}{17}},
  \bibinfo{pages}{311} (\bibinfo{year}{2021}).

\bibitem[{\citenamefont{Golde et~al.}(2008)\citenamefont{Golde, Meier, and
  Koch}}]{golde2008}
\bibinfo{author}{\bibfnamefont{D.}~\bibnamefont{Golde}},
  \bibinfo{author}{\bibfnamefont{T.}~\bibnamefont{Meier}}, \bibnamefont{and}
  \bibinfo{author}{\bibfnamefont{S.~W.} \bibnamefont{Koch}},
  \bibinfo{journal}{Physical Review B} \textbf{\bibinfo{volume}{77}},
  \bibinfo{pages}{075330} (\bibinfo{year}{2008}).

\bibitem[{\citenamefont{Vampa et~al.}(2014)\citenamefont{Vampa, McDonald,
  Orlando, Klug, Corkum, and Brabec}}]{vampa2014theoretical}
\bibinfo{author}{\bibfnamefont{G.}~\bibnamefont{Vampa}},
  \bibinfo{author}{\bibfnamefont{C.}~\bibnamefont{McDonald}},
  \bibinfo{author}{\bibfnamefont{G.}~\bibnamefont{Orlando}},
  \bibinfo{author}{\bibfnamefont{D.}~\bibnamefont{Klug}},
  \bibinfo{author}{\bibfnamefont{P.}~\bibnamefont{Corkum}}, \bibnamefont{and}
  \bibinfo{author}{\bibfnamefont{T.}~\bibnamefont{Brabec}},
  \bibinfo{journal}{Physical Review Letters} \textbf{\bibinfo{volume}{113}},
  \bibinfo{pages}{073901} (\bibinfo{year}{2014}).

\bibitem[{\citenamefont{Vampa et~al.}(2015{\natexlab{b}})\citenamefont{Vampa,
  McDonald, Orlando, Corkum, and Brabec}}]{vampa2015semiclassical}
\bibinfo{author}{\bibfnamefont{G.}~\bibnamefont{Vampa}},
  \bibinfo{author}{\bibfnamefont{C.}~\bibnamefont{McDonald}},
  \bibinfo{author}{\bibfnamefont{G.}~\bibnamefont{Orlando}},
  \bibinfo{author}{\bibfnamefont{P.}~\bibnamefont{Corkum}}, \bibnamefont{and}
  \bibinfo{author}{\bibfnamefont{T.}~\bibnamefont{Brabec}},
  \bibinfo{journal}{Physical Review B} \textbf{\bibinfo{volume}{91}},
  \bibinfo{pages}{064302} (\bibinfo{year}{2015}{\natexlab{b}}).

\bibitem[{\citenamefont{Ikemachi et~al.}(2017)\citenamefont{Ikemachi,
  Shinohara, Sato, Yumoto, Kuwata-Gonokami, and
  Ishikawa}}]{ikemachi2017trajectory}
\bibinfo{author}{\bibfnamefont{T.}~\bibnamefont{Ikemachi}},
  \bibinfo{author}{\bibfnamefont{Y.}~\bibnamefont{Shinohara}},
  \bibinfo{author}{\bibfnamefont{T.}~\bibnamefont{Sato}},
  \bibinfo{author}{\bibfnamefont{J.}~\bibnamefont{Yumoto}},
  \bibinfo{author}{\bibfnamefont{M.}~\bibnamefont{Kuwata-Gonokami}},
  \bibnamefont{and} \bibinfo{author}{\bibfnamefont{K.~L.}
  \bibnamefont{Ishikawa}}, \bibinfo{journal}{Physical Review A}
  \textbf{\bibinfo{volume}{95}}, \bibinfo{pages}{043416}
  (\bibinfo{year}{2017}).

\bibitem[{\citenamefont{Tancogne-Dejean
  et~al.}(2017{\natexlab{b}})\citenamefont{Tancogne-Dejean, M{\"u}cke,
  K{\"a}rtner, and Rubio}}]{tancogne2017impact}
\bibinfo{author}{\bibfnamefont{N.}~\bibnamefont{Tancogne-Dejean}},
  \bibinfo{author}{\bibfnamefont{O.~D.} \bibnamefont{M{\"u}cke}},
  \bibinfo{author}{\bibfnamefont{F.~X.} \bibnamefont{K{\"a}rtner}},
  \bibnamefont{and} \bibinfo{author}{\bibfnamefont{A.}~\bibnamefont{Rubio}},
  \bibinfo{journal}{Physical Review Letters} \textbf{\bibinfo{volume}{118}},
  \bibinfo{pages}{087403} (\bibinfo{year}{2017}{\natexlab{b}}).

\bibitem[{\citenamefont{Floss et~al.}(2018)\citenamefont{Floss, Lemell,
  Wachter, Smejkal, Sato, Tong, Yabana, and Burgd{\"o}rfer}}]{floss2018}
\bibinfo{author}{\bibfnamefont{I.}~\bibnamefont{Floss}},
  \bibinfo{author}{\bibfnamefont{C.}~\bibnamefont{Lemell}},
  \bibinfo{author}{\bibfnamefont{G.}~\bibnamefont{Wachter}},
  \bibinfo{author}{\bibfnamefont{V.}~\bibnamefont{Smejkal}},
  \bibinfo{author}{\bibfnamefont{S.~A.} \bibnamefont{Sato}},
  \bibinfo{author}{\bibfnamefont{X.-M.} \bibnamefont{Tong}},
  \bibinfo{author}{\bibfnamefont{K.}~\bibnamefont{Yabana}}, \bibnamefont{and}
  \bibinfo{author}{\bibfnamefont{J.}~\bibnamefont{Burgd{\"o}rfer}},
  \bibinfo{journal}{Physical Review A} \textbf{\bibinfo{volume}{97}},
  \bibinfo{pages}{011401} (\bibinfo{year}{2018}).

\bibitem[{\citenamefont{Sato et~al.}(2021)\citenamefont{Sato, Hirori, Sanari,
  Kanemitsu, and Rubio}}]{sato2021high}
\bibinfo{author}{\bibfnamefont{S.~A.} \bibnamefont{Sato}},
  \bibinfo{author}{\bibfnamefont{H.}~\bibnamefont{Hirori}},
  \bibinfo{author}{\bibfnamefont{Y.}~\bibnamefont{Sanari}},
  \bibinfo{author}{\bibfnamefont{Y.}~\bibnamefont{Kanemitsu}},
  \bibnamefont{and} \bibinfo{author}{\bibfnamefont{A.}~\bibnamefont{Rubio}},
  \bibinfo{journal}{Physical Review B} \textbf{\bibinfo{volume}{103}},
  \bibinfo{pages}{L041408} (\bibinfo{year}{2021}).

\bibitem[{\citenamefont{Armitage et~al.}(2018)\citenamefont{Armitage, Mele, and
  Vishwanath}}]{armitage2018}
\bibinfo{author}{\bibfnamefont{N.}~\bibnamefont{Armitage}},
  \bibinfo{author}{\bibfnamefont{E.}~\bibnamefont{Mele}}, \bibnamefont{and}
  \bibinfo{author}{\bibfnamefont{A.}~\bibnamefont{Vishwanath}},
  \bibinfo{journal}{Reviews of Modern Physics} \textbf{\bibinfo{volume}{90}},
  \bibinfo{pages}{015001} (\bibinfo{year}{2018}).

\bibitem[{\citenamefont{Hu et~al.}(2019)\citenamefont{Hu, Xu, Ni, and
  Mao}}]{hu2019}
\bibinfo{author}{\bibfnamefont{J.}~\bibnamefont{Hu}},
  \bibinfo{author}{\bibfnamefont{S.-Y.} \bibnamefont{Xu}},
  \bibinfo{author}{\bibfnamefont{N.}~\bibnamefont{Ni}}, \bibnamefont{and}
  \bibinfo{author}{\bibfnamefont{Z.}~\bibnamefont{Mao}},
  \bibinfo{journal}{Annual Review of Materials Research}
  \textbf{\bibinfo{volume}{49}}, \bibinfo{pages}{207} (\bibinfo{year}{2019}).

\bibitem[{\citenamefont{Mics et~al.}(2015)\citenamefont{Mics, Tielrooij,
  Parvez, Jensen, Ivanov, Feng, M{\"u}llen, Bonn, and Turchinovich}}]{mics2015}
\bibinfo{author}{\bibfnamefont{Z.}~\bibnamefont{Mics}},
  \bibinfo{author}{\bibfnamefont{K.-J.} \bibnamefont{Tielrooij}},
  \bibinfo{author}{\bibfnamefont{K.}~\bibnamefont{Parvez}},
  \bibinfo{author}{\bibfnamefont{S.~A.} \bibnamefont{Jensen}},
  \bibinfo{author}{\bibfnamefont{I.}~\bibnamefont{Ivanov}},
  \bibinfo{author}{\bibfnamefont{X.}~\bibnamefont{Feng}},
  \bibinfo{author}{\bibfnamefont{K.}~\bibnamefont{M{\"u}llen}},
  \bibinfo{author}{\bibfnamefont{M.}~\bibnamefont{Bonn}}, \bibnamefont{and}
  \bibinfo{author}{\bibfnamefont{D.}~\bibnamefont{Turchinovich}},
  \bibinfo{journal}{Nature communications} \textbf{\bibinfo{volume}{6}},
  \bibinfo{pages}{1} (\bibinfo{year}{2015}).

\bibitem[{\citenamefont{Yoshikawa et~al.}(2017)\citenamefont{Yoshikawa, Tamaya,
  and Tanaka}}]{yoshikawa2017}
\bibinfo{author}{\bibfnamefont{N.}~\bibnamefont{Yoshikawa}},
  \bibinfo{author}{\bibfnamefont{T.}~\bibnamefont{Tamaya}}, \bibnamefont{and}
  \bibinfo{author}{\bibfnamefont{K.}~\bibnamefont{Tanaka}},
  \bibinfo{journal}{Science} \textbf{\bibinfo{volume}{356}},
  \bibinfo{pages}{736} (\bibinfo{year}{2017}).

\bibitem[{\citenamefont{Hafez et~al.}(2018)\citenamefont{Hafez, Kovalev,
  Deinert, Mics, Green, Awari, Chen, Germanskiy, Lehnert, Teichert
  et~al.}}]{hafez2018}
\bibinfo{author}{\bibfnamefont{H.~A.} \bibnamefont{Hafez}},
  \bibinfo{author}{\bibfnamefont{S.}~\bibnamefont{Kovalev}},
  \bibinfo{author}{\bibfnamefont{J.-C.} \bibnamefont{Deinert}},
  \bibinfo{author}{\bibfnamefont{Z.}~\bibnamefont{Mics}},
  \bibinfo{author}{\bibfnamefont{B.}~\bibnamefont{Green}},
  \bibinfo{author}{\bibfnamefont{N.}~\bibnamefont{Awari}},
  \bibinfo{author}{\bibfnamefont{M.}~\bibnamefont{Chen}},
  \bibinfo{author}{\bibfnamefont{S.}~\bibnamefont{Germanskiy}},
  \bibinfo{author}{\bibfnamefont{U.}~\bibnamefont{Lehnert}},
  \bibinfo{author}{\bibfnamefont{J.}~\bibnamefont{Teichert}},
  \bibnamefont{et~al.}, \bibinfo{journal}{Nature}
  \textbf{\bibinfo{volume}{561}}, \bibinfo{pages}{507} (\bibinfo{year}{2018}).

\bibitem[{\citenamefont{Shan et~al.}(2018)\citenamefont{Shan, Li, Huang, Tong,
  Yao, Liu, and Wu}}]{shan2018}
\bibinfo{author}{\bibfnamefont{Y.}~\bibnamefont{Shan}},
  \bibinfo{author}{\bibfnamefont{Y.}~\bibnamefont{Li}},
  \bibinfo{author}{\bibfnamefont{D.}~\bibnamefont{Huang}},
  \bibinfo{author}{\bibfnamefont{Q.}~\bibnamefont{Tong}},
  \bibinfo{author}{\bibfnamefont{W.}~\bibnamefont{Yao}},
  \bibinfo{author}{\bibfnamefont{W.-T.} \bibnamefont{Liu}}, \bibnamefont{and}
  \bibinfo{author}{\bibfnamefont{S.}~\bibnamefont{Wu}},
  \bibinfo{journal}{Science advances} \textbf{\bibinfo{volume}{4}},
  \bibinfo{pages}{eaat0074} (\bibinfo{year}{2018}).

\bibitem[{\citenamefont{Tancogne-Dejean and Rubio}(2018)}]{tancogne2018}
\bibinfo{author}{\bibfnamefont{N.}~\bibnamefont{Tancogne-Dejean}}
  \bibnamefont{and} \bibinfo{author}{\bibfnamefont{A.}~\bibnamefont{Rubio}},
  \bibinfo{journal}{Science advances} \textbf{\bibinfo{volume}{4}},
  \bibinfo{pages}{eaao5207} (\bibinfo{year}{2018}).

\bibitem[{\citenamefont{Ikeda}(2020)}]{ikeda2020high}
\bibinfo{author}{\bibfnamefont{T.~N.} \bibnamefont{Ikeda}},
  \bibinfo{journal}{Physical Review Research} \textbf{\bibinfo{volume}{2}},
  \bibinfo{pages}{032015} (\bibinfo{year}{2020}).

\bibitem[{\citenamefont{Alonso~Calafell
  et~al.}(2021)\citenamefont{Alonso~Calafell, Rozema, Alcaraz~Iranzo, Trenti,
  Jenke, Cox, Kumar, Bieliaiev, Nanot, Peng et~al.}}]{alonso2021giant}
\bibinfo{author}{\bibfnamefont{I.}~\bibnamefont{Alonso~Calafell}},
  \bibinfo{author}{\bibfnamefont{L.~A.} \bibnamefont{Rozema}},
  \bibinfo{author}{\bibfnamefont{D.}~\bibnamefont{Alcaraz~Iranzo}},
  \bibinfo{author}{\bibfnamefont{A.}~\bibnamefont{Trenti}},
  \bibinfo{author}{\bibfnamefont{P.~K.} \bibnamefont{Jenke}},
  \bibinfo{author}{\bibfnamefont{J.~D.} \bibnamefont{Cox}},
  \bibinfo{author}{\bibfnamefont{A.}~\bibnamefont{Kumar}},
  \bibinfo{author}{\bibfnamefont{H.}~\bibnamefont{Bieliaiev}},
  \bibinfo{author}{\bibfnamefont{S.}~\bibnamefont{Nanot}},
  \bibinfo{author}{\bibfnamefont{C.}~\bibnamefont{Peng}}, \bibnamefont{et~al.},
  \bibinfo{journal}{Nature Nanotechnology} \textbf{\bibinfo{volume}{16}},
  \bibinfo{pages}{318} (\bibinfo{year}{2021}).

\bibitem[{\citenamefont{Cheng et~al.}(2020)\citenamefont{Cheng, Kanda, Ikeda,
  Matsuda, Xia, Schumann, Stemmer, Itatani, Armitage, and
  Matsunaga}}]{cheng2020}
\bibinfo{author}{\bibfnamefont{B.}~\bibnamefont{Cheng}},
  \bibinfo{author}{\bibfnamefont{N.}~\bibnamefont{Kanda}},
  \bibinfo{author}{\bibfnamefont{T.~N.} \bibnamefont{Ikeda}},
  \bibinfo{author}{\bibfnamefont{T.}~\bibnamefont{Matsuda}},
  \bibinfo{author}{\bibfnamefont{P.}~\bibnamefont{Xia}},
  \bibinfo{author}{\bibfnamefont{T.}~\bibnamefont{Schumann}},
  \bibinfo{author}{\bibfnamefont{S.}~\bibnamefont{Stemmer}},
  \bibinfo{author}{\bibfnamefont{J.}~\bibnamefont{Itatani}},
  \bibinfo{author}{\bibfnamefont{N.}~\bibnamefont{Armitage}}, \bibnamefont{and}
  \bibinfo{author}{\bibfnamefont{R.}~\bibnamefont{Matsunaga}},
  \bibinfo{journal}{Physical Review Letters} \textbf{\bibinfo{volume}{124}},
  \bibinfo{pages}{117402} (\bibinfo{year}{2020}).

\bibitem[{\citenamefont{Huang et~al.}(2015)\citenamefont{Huang, Xu, Belopolski,
  Lee, Chang, Wang, Alidoust, Bian, Neupane, Zhang et~al.}}]{huang2015}
\bibinfo{author}{\bibfnamefont{S.-M.} \bibnamefont{Huang}},
  \bibinfo{author}{\bibfnamefont{S.-Y.} \bibnamefont{Xu}},
  \bibinfo{author}{\bibfnamefont{I.}~\bibnamefont{Belopolski}},
  \bibinfo{author}{\bibfnamefont{C.-C.} \bibnamefont{Lee}},
  \bibinfo{author}{\bibfnamefont{G.}~\bibnamefont{Chang}},
  \bibinfo{author}{\bibfnamefont{B.}~\bibnamefont{Wang}},
  \bibinfo{author}{\bibfnamefont{N.}~\bibnamefont{Alidoust}},
  \bibinfo{author}{\bibfnamefont{G.}~\bibnamefont{Bian}},
  \bibinfo{author}{\bibfnamefont{M.}~\bibnamefont{Neupane}},
  \bibinfo{author}{\bibfnamefont{C.}~\bibnamefont{Zhang}},
  \bibnamefont{et~al.}, \bibinfo{journal}{Nature communications}
  \textbf{\bibinfo{volume}{6}}, \bibinfo{pages}{1} (\bibinfo{year}{2015}).

\bibitem[{\citenamefont{Weng et~al.}(2015)\citenamefont{Weng, Fang, Fang,
  Bernevig, and Dai}}]{weng2015}
\bibinfo{author}{\bibfnamefont{H.}~\bibnamefont{Weng}},
  \bibinfo{author}{\bibfnamefont{C.}~\bibnamefont{Fang}},
  \bibinfo{author}{\bibfnamefont{Z.}~\bibnamefont{Fang}},
  \bibinfo{author}{\bibfnamefont{B.~A.} \bibnamefont{Bernevig}},
  \bibnamefont{and} \bibinfo{author}{\bibfnamefont{X.}~\bibnamefont{Dai}},
  \bibinfo{journal}{Physical Review X} \textbf{\bibinfo{volume}{5}},
  \bibinfo{pages}{011029} (\bibinfo{year}{2015}).

\bibitem[{\citenamefont{Murakami et~al.}(2017)\citenamefont{Murakami, Hirayama,
  Okugawa, and Miyake}}]{murakami2017}
\bibinfo{author}{\bibfnamefont{S.}~\bibnamefont{Murakami}},
  \bibinfo{author}{\bibfnamefont{M.}~\bibnamefont{Hirayama}},
  \bibinfo{author}{\bibfnamefont{R.}~\bibnamefont{Okugawa}}, \bibnamefont{and}
  \bibinfo{author}{\bibfnamefont{T.}~\bibnamefont{Miyake}},
  \bibinfo{journal}{Science advances} \textbf{\bibinfo{volume}{3}},
  \bibinfo{pages}{e1602680} (\bibinfo{year}{2017}).

\bibitem[{\citenamefont{Lv et~al.}(2015{\natexlab{a}})\citenamefont{Lv, Weng,
  Fu, Wang, Miao, Ma, Richard, Huang, Zhao, Chen et~al.}}]{lv2015}
\bibinfo{author}{\bibfnamefont{B.}~\bibnamefont{Lv}},
  \bibinfo{author}{\bibfnamefont{H.}~\bibnamefont{Weng}},
  \bibinfo{author}{\bibfnamefont{B.}~\bibnamefont{Fu}},
  \bibinfo{author}{\bibfnamefont{X.~P.} \bibnamefont{Wang}},
  \bibinfo{author}{\bibfnamefont{H.}~\bibnamefont{Miao}},
  \bibinfo{author}{\bibfnamefont{J.}~\bibnamefont{Ma}},
  \bibinfo{author}{\bibfnamefont{P.}~\bibnamefont{Richard}},
  \bibinfo{author}{\bibfnamefont{X.}~\bibnamefont{Huang}},
  \bibinfo{author}{\bibfnamefont{L.}~\bibnamefont{Zhao}},
  \bibinfo{author}{\bibfnamefont{G.}~\bibnamefont{Chen}}, \bibnamefont{et~al.},
  \bibinfo{journal}{Physical Review X} \textbf{\bibinfo{volume}{5}},
  \bibinfo{pages}{031013} (\bibinfo{year}{2015}{\natexlab{a}}).

\bibitem[{\citenamefont{Lv et~al.}(2015{\natexlab{b}})\citenamefont{Lv, Xu,
  Weng, Ma, Richard, Huang, Zhao, Chen, Matt, Bisti et~al.}}]{lv2015observ}
\bibinfo{author}{\bibfnamefont{B.}~\bibnamefont{Lv}},
  \bibinfo{author}{\bibfnamefont{N.}~\bibnamefont{Xu}},
  \bibinfo{author}{\bibfnamefont{H.}~\bibnamefont{Weng}},
  \bibinfo{author}{\bibfnamefont{J.}~\bibnamefont{Ma}},
  \bibinfo{author}{\bibfnamefont{P.}~\bibnamefont{Richard}},
  \bibinfo{author}{\bibfnamefont{X.}~\bibnamefont{Huang}},
  \bibinfo{author}{\bibfnamefont{L.}~\bibnamefont{Zhao}},
  \bibinfo{author}{\bibfnamefont{G.}~\bibnamefont{Chen}},
  \bibinfo{author}{\bibfnamefont{C.}~\bibnamefont{Matt}},
  \bibinfo{author}{\bibfnamefont{F.}~\bibnamefont{Bisti}},
  \bibnamefont{et~al.}, \bibinfo{journal}{Nature Physics}
  \textbf{\bibinfo{volume}{11}}, \bibinfo{pages}{724}
  (\bibinfo{year}{2015}{\natexlab{b}}).

\bibitem[{\citenamefont{Xu et~al.}(2015{\natexlab{a}})\citenamefont{Xu,
  Belopolski, Sanchez, Zhang, Chang, Guo, Bian, Yuan, Lu, Chang
  et~al.}}]{xu2015}
\bibinfo{author}{\bibfnamefont{S.-Y.} \bibnamefont{Xu}},
  \bibinfo{author}{\bibfnamefont{I.}~\bibnamefont{Belopolski}},
  \bibinfo{author}{\bibfnamefont{D.~S.} \bibnamefont{Sanchez}},
  \bibinfo{author}{\bibfnamefont{C.}~\bibnamefont{Zhang}},
  \bibinfo{author}{\bibfnamefont{G.}~\bibnamefont{Chang}},
  \bibinfo{author}{\bibfnamefont{C.}~\bibnamefont{Guo}},
  \bibinfo{author}{\bibfnamefont{G.}~\bibnamefont{Bian}},
  \bibinfo{author}{\bibfnamefont{Z.}~\bibnamefont{Yuan}},
  \bibinfo{author}{\bibfnamefont{H.}~\bibnamefont{Lu}},
  \bibinfo{author}{\bibfnamefont{T.-R.} \bibnamefont{Chang}},
  \bibnamefont{et~al.}, \bibinfo{journal}{Science advances}
  \textbf{\bibinfo{volume}{1}}, \bibinfo{pages}{e1501092}
  (\bibinfo{year}{2015}{\natexlab{a}}).

\bibitem[{\citenamefont{Arnold et~al.}(2016)\citenamefont{Arnold, Naumann, Wu,
  Sun, Schmidt, Borrmann, Felser, Yan, and Hassinger}}]{arnold2016}
\bibinfo{author}{\bibfnamefont{F.}~\bibnamefont{Arnold}},
  \bibinfo{author}{\bibfnamefont{M.}~\bibnamefont{Naumann}},
  \bibinfo{author}{\bibfnamefont{S.-C.} \bibnamefont{Wu}},
  \bibinfo{author}{\bibfnamefont{Y.}~\bibnamefont{Sun}},
  \bibinfo{author}{\bibfnamefont{M.}~\bibnamefont{Schmidt}},
  \bibinfo{author}{\bibfnamefont{H.}~\bibnamefont{Borrmann}},
  \bibinfo{author}{\bibfnamefont{C.}~\bibnamefont{Felser}},
  \bibinfo{author}{\bibfnamefont{B.}~\bibnamefont{Yan}}, \bibnamefont{and}
  \bibinfo{author}{\bibfnamefont{E.}~\bibnamefont{Hassinger}},
  \bibinfo{journal}{Physical review letters} \textbf{\bibinfo{volume}{117}},
  \bibinfo{pages}{146401} (\bibinfo{year}{2016}).

\bibitem[{\citenamefont{Xu et~al.}(2015{\natexlab{b}})\citenamefont{Xu,
  Belopolski, Alidoust, Neupane, Bian, Zhang, Sankar, Chang, Yuan, Lee
  et~al.}}]{xu2015discovery}
\bibinfo{author}{\bibfnamefont{S.-Y.} \bibnamefont{Xu}},
  \bibinfo{author}{\bibfnamefont{I.}~\bibnamefont{Belopolski}},
  \bibinfo{author}{\bibfnamefont{N.}~\bibnamefont{Alidoust}},
  \bibinfo{author}{\bibfnamefont{M.}~\bibnamefont{Neupane}},
  \bibinfo{author}{\bibfnamefont{G.}~\bibnamefont{Bian}},
  \bibinfo{author}{\bibfnamefont{C.}~\bibnamefont{Zhang}},
  \bibinfo{author}{\bibfnamefont{R.}~\bibnamefont{Sankar}},
  \bibinfo{author}{\bibfnamefont{G.}~\bibnamefont{Chang}},
  \bibinfo{author}{\bibfnamefont{Z.}~\bibnamefont{Yuan}},
  \bibinfo{author}{\bibfnamefont{C.-C.} \bibnamefont{Lee}},
  \bibnamefont{et~al.}, \bibinfo{journal}{Science}
  \textbf{\bibinfo{volume}{349}}, \bibinfo{pages}{613}
  (\bibinfo{year}{2015}{\natexlab{b}}).

\bibitem[{\citenamefont{Xu et~al.}(2016)\citenamefont{Xu, Weng, Lv, Matt, Park,
  Bisti, Strocov, Gawryluk, Pomjakushina, Conder et~al.}}]{xu2016observation}
\bibinfo{author}{\bibfnamefont{N.}~\bibnamefont{Xu}},
  \bibinfo{author}{\bibfnamefont{H.}~\bibnamefont{Weng}},
  \bibinfo{author}{\bibfnamefont{B.}~\bibnamefont{Lv}},
  \bibinfo{author}{\bibfnamefont{C.~E.} \bibnamefont{Matt}},
  \bibinfo{author}{\bibfnamefont{J.}~\bibnamefont{Park}},
  \bibinfo{author}{\bibfnamefont{F.}~\bibnamefont{Bisti}},
  \bibinfo{author}{\bibfnamefont{V.~N.} \bibnamefont{Strocov}},
  \bibinfo{author}{\bibfnamefont{D.}~\bibnamefont{Gawryluk}},
  \bibinfo{author}{\bibfnamefont{E.}~\bibnamefont{Pomjakushina}},
  \bibinfo{author}{\bibfnamefont{K.}~\bibnamefont{Conder}},
  \bibnamefont{et~al.}, \bibinfo{journal}{Nature communications}
  \textbf{\bibinfo{volume}{7}}, \bibinfo{pages}{1} (\bibinfo{year}{2016}).

\bibitem[{\citenamefont{Soluyanov et~al.}(2015)\citenamefont{Soluyanov, Gresch,
  Wang, Wu, Troyer, Dai, and Bernevig}}]{soluyanov2015}
\bibinfo{author}{\bibfnamefont{A.~A.} \bibnamefont{Soluyanov}},
  \bibinfo{author}{\bibfnamefont{D.}~\bibnamefont{Gresch}},
  \bibinfo{author}{\bibfnamefont{Z.}~\bibnamefont{Wang}},
  \bibinfo{author}{\bibfnamefont{Q.}~\bibnamefont{Wu}},
  \bibinfo{author}{\bibfnamefont{M.}~\bibnamefont{Troyer}},
  \bibinfo{author}{\bibfnamefont{X.}~\bibnamefont{Dai}}, \bibnamefont{and}
  \bibinfo{author}{\bibfnamefont{B.~A.} \bibnamefont{Bernevig}},
  \bibinfo{journal}{Nature} \textbf{\bibinfo{volume}{527}},
  \bibinfo{pages}{495} (\bibinfo{year}{2015}).

\bibitem[{\citenamefont{Sodemann and Fu}(2015)}]{sodemann2015}
\bibinfo{author}{\bibfnamefont{I.}~\bibnamefont{Sodemann}} \bibnamefont{and}
  \bibinfo{author}{\bibfnamefont{L.}~\bibnamefont{Fu}},
  \bibinfo{journal}{Physical review letters} \textbf{\bibinfo{volume}{115}},
  \bibinfo{pages}{216806} (\bibinfo{year}{2015}).

\bibitem[{\citenamefont{Zhang et~al.}(2018)\citenamefont{Zhang, Sun, and
  Yan}}]{zhang2018}
\bibinfo{author}{\bibfnamefont{Y.}~\bibnamefont{Zhang}},
  \bibinfo{author}{\bibfnamefont{Y.}~\bibnamefont{Sun}}, \bibnamefont{and}
  \bibinfo{author}{\bibfnamefont{B.}~\bibnamefont{Yan}},
  \bibinfo{journal}{Physical Review B} \textbf{\bibinfo{volume}{97}},
  \bibinfo{pages}{041101} (\bibinfo{year}{2018}).

\bibitem[{\citenamefont{Rostami and Polini}(2018)}]{rostami2018}
\bibinfo{author}{\bibfnamefont{H.}~\bibnamefont{Rostami}} \bibnamefont{and}
  \bibinfo{author}{\bibfnamefont{M.}~\bibnamefont{Polini}},
  \bibinfo{journal}{Physical Review B} \textbf{\bibinfo{volume}{97}},
  \bibinfo{pages}{195151} (\bibinfo{year}{2018}).

\bibitem[{\citenamefont{Kang et~al.}(2019)\citenamefont{Kang, Li, Sohn, Shan,
  and Mak}}]{kang2019}
\bibinfo{author}{\bibfnamefont{K.}~\bibnamefont{Kang}},
  \bibinfo{author}{\bibfnamefont{T.}~\bibnamefont{Li}},
  \bibinfo{author}{\bibfnamefont{E.}~\bibnamefont{Sohn}},
  \bibinfo{author}{\bibfnamefont{J.}~\bibnamefont{Shan}}, \bibnamefont{and}
  \bibinfo{author}{\bibfnamefont{K.~F.} \bibnamefont{Mak}},
  \bibinfo{journal}{Nature materials} \textbf{\bibinfo{volume}{18}},
  \bibinfo{pages}{324} (\bibinfo{year}{2019}).

\bibitem[{\citenamefont{Wawrzik et~al.}(2021)\citenamefont{Wawrzik, You, Facio,
  Van Den~Brink, and Sodemann}}]{wawrzik2021}
\bibinfo{author}{\bibfnamefont{D.}~\bibnamefont{Wawrzik}},
  \bibinfo{author}{\bibfnamefont{J.-S.} \bibnamefont{You}},
  \bibinfo{author}{\bibfnamefont{J.~I.} \bibnamefont{Facio}},
  \bibinfo{author}{\bibfnamefont{J.}~\bibnamefont{Van Den~Brink}},
  \bibnamefont{and} \bibinfo{author}{\bibfnamefont{I.}~\bibnamefont{Sodemann}},
  \bibinfo{journal}{Physical Review Letters} \textbf{\bibinfo{volume}{127}},
  \bibinfo{pages}{056601} (\bibinfo{year}{2021}).

\bibitem[{\citenamefont{Chan et~al.}(2017)\citenamefont{Chan, Lindner, Refael,
  and Lee}}]{chan2017}
\bibinfo{author}{\bibfnamefont{C.-K.} \bibnamefont{Chan}},
  \bibinfo{author}{\bibfnamefont{N.~H.} \bibnamefont{Lindner}},
  \bibinfo{author}{\bibfnamefont{G.}~\bibnamefont{Refael}}, \bibnamefont{and}
  \bibinfo{author}{\bibfnamefont{P.~A.} \bibnamefont{Lee}},
  \bibinfo{journal}{Physical Review B} \textbf{\bibinfo{volume}{95}},
  \bibinfo{pages}{041104} (\bibinfo{year}{2017}).

\bibitem[{\citenamefont{Sirica et~al.}(2019)\citenamefont{Sirica, Tobey, Zhao,
  Chen, Xu, Yang, Shen, Yarotski, Bowlan, Trugman et~al.}}]{sirica2019}
\bibinfo{author}{\bibfnamefont{N.}~\bibnamefont{Sirica}},
  \bibinfo{author}{\bibfnamefont{R.}~\bibnamefont{Tobey}},
  \bibinfo{author}{\bibfnamefont{L.}~\bibnamefont{Zhao}},
  \bibinfo{author}{\bibfnamefont{G.}~\bibnamefont{Chen}},
  \bibinfo{author}{\bibfnamefont{B.}~\bibnamefont{Xu}},
  \bibinfo{author}{\bibfnamefont{R.}~\bibnamefont{Yang}},
  \bibinfo{author}{\bibfnamefont{B.}~\bibnamefont{Shen}},
  \bibinfo{author}{\bibfnamefont{D.}~\bibnamefont{Yarotski}},
  \bibinfo{author}{\bibfnamefont{P.}~\bibnamefont{Bowlan}},
  \bibinfo{author}{\bibfnamefont{S.}~\bibnamefont{Trugman}},
  \bibnamefont{et~al.}, \bibinfo{journal}{Physical review letters}
  \textbf{\bibinfo{volume}{122}}, \bibinfo{pages}{197401}
  (\bibinfo{year}{2019}).

\bibitem[{\citenamefont{Sirica et~al.}(2022)\citenamefont{Sirica, Orth,
  Scheurer, Dai, Lee, Padmanabhan, Mix, Teitelbaum, Trigo, Zhao
  et~al.}}]{sirica2022}
\bibinfo{author}{\bibfnamefont{N.}~\bibnamefont{Sirica}},
  \bibinfo{author}{\bibfnamefont{P.~P.} \bibnamefont{Orth}},
  \bibinfo{author}{\bibfnamefont{M.}~\bibnamefont{Scheurer}},
  \bibinfo{author}{\bibfnamefont{Y.}~\bibnamefont{Dai}},
  \bibinfo{author}{\bibfnamefont{M.-C.} \bibnamefont{Lee}},
  \bibinfo{author}{\bibfnamefont{P.}~\bibnamefont{Padmanabhan}},
  \bibinfo{author}{\bibfnamefont{L.}~\bibnamefont{Mix}},
  \bibinfo{author}{\bibfnamefont{S.}~\bibnamefont{Teitelbaum}},
  \bibinfo{author}{\bibfnamefont{M.}~\bibnamefont{Trigo}},
  \bibinfo{author}{\bibfnamefont{L.}~\bibnamefont{Zhao}}, \bibnamefont{et~al.},
  \bibinfo{journal}{Nature materials} \textbf{\bibinfo{volume}{21}},
  \bibinfo{pages}{62} (\bibinfo{year}{2022}).

\bibitem[{\citenamefont{de~Juan et~al.}(2017)\citenamefont{de~Juan, Grushin,
  Morimoto, and Moore}}]{de2017}
\bibinfo{author}{\bibfnamefont{F.}~\bibnamefont{de~Juan}},
  \bibinfo{author}{\bibfnamefont{A.~G.} \bibnamefont{Grushin}},
  \bibinfo{author}{\bibfnamefont{T.}~\bibnamefont{Morimoto}}, \bibnamefont{and}
  \bibinfo{author}{\bibfnamefont{J.~E.} \bibnamefont{Moore}},
  \bibinfo{journal}{Nature communications} \textbf{\bibinfo{volume}{8}},
  \bibinfo{pages}{1} (\bibinfo{year}{2017}).

\bibitem[{\citenamefont{Wang et~al.}(2019)\citenamefont{Wang, Zheng, He, Cao,
  Liu, Wang, Ma, Lai, Lu, Jia et~al.}}]{wang2019}
\bibinfo{author}{\bibfnamefont{Q.}~\bibnamefont{Wang}},
  \bibinfo{author}{\bibfnamefont{J.}~\bibnamefont{Zheng}},
  \bibinfo{author}{\bibfnamefont{Y.}~\bibnamefont{He}},
  \bibinfo{author}{\bibfnamefont{J.}~\bibnamefont{Cao}},
  \bibinfo{author}{\bibfnamefont{X.}~\bibnamefont{Liu}},
  \bibinfo{author}{\bibfnamefont{M.}~\bibnamefont{Wang}},
  \bibinfo{author}{\bibfnamefont{J.}~\bibnamefont{Ma}},
  \bibinfo{author}{\bibfnamefont{J.}~\bibnamefont{Lai}},
  \bibinfo{author}{\bibfnamefont{H.}~\bibnamefont{Lu}},
  \bibinfo{author}{\bibfnamefont{S.}~\bibnamefont{Jia}}, \bibnamefont{et~al.},
  \bibinfo{journal}{Nature communications} \textbf{\bibinfo{volume}{10}},
  \bibinfo{pages}{1} (\bibinfo{year}{2019}).

\bibitem[{\citenamefont{Ma et~al.}(2019)\citenamefont{Ma, Gu, Liu, Lai, Yu,
  Zhuo, Liu, Chen, Feng, and Sun}}]{ma2019}
\bibinfo{author}{\bibfnamefont{J.}~\bibnamefont{Ma}},
  \bibinfo{author}{\bibfnamefont{Q.}~\bibnamefont{Gu}},
  \bibinfo{author}{\bibfnamefont{Y.}~\bibnamefont{Liu}},
  \bibinfo{author}{\bibfnamefont{J.}~\bibnamefont{Lai}},
  \bibinfo{author}{\bibfnamefont{P.}~\bibnamefont{Yu}},
  \bibinfo{author}{\bibfnamefont{X.}~\bibnamefont{Zhuo}},
  \bibinfo{author}{\bibfnamefont{Z.}~\bibnamefont{Liu}},
  \bibinfo{author}{\bibfnamefont{J.-H.} \bibnamefont{Chen}},
  \bibinfo{author}{\bibfnamefont{J.}~\bibnamefont{Feng}}, \bibnamefont{and}
  \bibinfo{author}{\bibfnamefont{D.}~\bibnamefont{Sun}},
  \bibinfo{journal}{Nature materials} \textbf{\bibinfo{volume}{18}},
  \bibinfo{pages}{476} (\bibinfo{year}{2019}).

\bibitem[{\citenamefont{Lv et~al.}(2021)\citenamefont{Lv, Xu, Han, Zhang, Han,
  Zhou, Yao, Liu, Lu, Weng et~al.}}]{lv2021}
\bibinfo{author}{\bibfnamefont{Y.-Y.} \bibnamefont{Lv}},
  \bibinfo{author}{\bibfnamefont{J.}~\bibnamefont{Xu}},
  \bibinfo{author}{\bibfnamefont{S.}~\bibnamefont{Han}},
  \bibinfo{author}{\bibfnamefont{C.}~\bibnamefont{Zhang}},
  \bibinfo{author}{\bibfnamefont{Y.}~\bibnamefont{Han}},
  \bibinfo{author}{\bibfnamefont{J.}~\bibnamefont{Zhou}},
  \bibinfo{author}{\bibfnamefont{S.-H.} \bibnamefont{Yao}},
  \bibinfo{author}{\bibfnamefont{X.-P.} \bibnamefont{Liu}},
  \bibinfo{author}{\bibfnamefont{M.-H.} \bibnamefont{Lu}},
  \bibinfo{author}{\bibfnamefont{H.}~\bibnamefont{Weng}}, \bibnamefont{et~al.},
  \bibinfo{journal}{Nature communications} \textbf{\bibinfo{volume}{12}},
  \bibinfo{pages}{1} (\bibinfo{year}{2021}).

\bibitem[{\citenamefont{Ishikawa}(2010)}]{ishikawa2010}
\bibinfo{author}{\bibfnamefont{K.~L.} \bibnamefont{Ishikawa}},
  \bibinfo{journal}{Physical Review B} \textbf{\bibinfo{volume}{82}},
  \bibinfo{pages}{201402} (\bibinfo{year}{2010}).

\bibitem[{\citenamefont{Lim et~al.}(2020)\citenamefont{Lim, Ang, de~Abajo,
  Kaminer, Ang, and Wong}}]{lim2020}
\bibinfo{author}{\bibfnamefont{J.}~\bibnamefont{Lim}},
  \bibinfo{author}{\bibfnamefont{Y.~S.} \bibnamefont{Ang}},
  \bibinfo{author}{\bibfnamefont{F.~J.~G.} \bibnamefont{de~Abajo}},
  \bibinfo{author}{\bibfnamefont{I.}~\bibnamefont{Kaminer}},
  \bibinfo{author}{\bibfnamefont{L.~K.} \bibnamefont{Ang}}, \bibnamefont{and}
  \bibinfo{author}{\bibfnamefont{L.~J.} \bibnamefont{Wong}},
  \bibinfo{journal}{Physical Review Research} \textbf{\bibinfo{volume}{2}},
  \bibinfo{pages}{043252} (\bibinfo{year}{2020}).

\bibitem[{\citenamefont{Carbotte}(2016)}]{carbotte2016}
\bibinfo{author}{\bibfnamefont{J.}~\bibnamefont{Carbotte}},
  \bibinfo{journal}{Physical Review B} \textbf{\bibinfo{volume}{94}},
  \bibinfo{pages}{165111} (\bibinfo{year}{2016}).

\bibitem[{\citenamefont{Neufeld et~al.}(2019)\citenamefont{Neufeld, Podolsky,
  and Cohen}}]{neufeld2019}
\bibinfo{author}{\bibfnamefont{O.}~\bibnamefont{Neufeld}},
  \bibinfo{author}{\bibfnamefont{D.}~\bibnamefont{Podolsky}}, \bibnamefont{and}
  \bibinfo{author}{\bibfnamefont{O.}~\bibnamefont{Cohen}},
  \bibinfo{journal}{Nature communications} \textbf{\bibinfo{volume}{10}},
  \bibinfo{pages}{1} (\bibinfo{year}{2019}).

\bibitem[{\citenamefont{McCormick et~al.}(2017)\citenamefont{McCormick, Kimchi,
  and Trivedi}}]{mccormick2017}
\bibinfo{author}{\bibfnamefont{T.~M.} \bibnamefont{McCormick}},
  \bibinfo{author}{\bibfnamefont{I.}~\bibnamefont{Kimchi}}, \bibnamefont{and}
  \bibinfo{author}{\bibfnamefont{N.}~\bibnamefont{Trivedi}},
  \bibinfo{journal}{Physical Review B} \textbf{\bibinfo{volume}{95}},
  \bibinfo{pages}{075133} (\bibinfo{year}{2017}).

\end{thebibliography}

\end{document}